\documentclass[journal]{IEEEtran}
\usepackage{cite}
\usepackage{lscape}
\usepackage{textcase}

\usepackage{amsfonts,amssymb,amsmath,bbm,mathtools,stackrel,epsfig,amsthm}
\usepackage[inline]{enumitem}
\usepackage[hidelinks]{hyperref}

\usepackage{scalerel}
\usepackage{tikz}
\usetikzlibrary{svg.path}

\definecolor{orcidlogocol}{HTML}{A6CE39}
\tikzset{
	orcidlogo/.pic={
		\fill[orcidlogocol] svg{M256,128c0,70.7-57.3,128-128,128C57.3,256,0,198.7,0,128C0,57.3,57.3,0,128,0C198.7,0,256,57.3,256,128z};
		\fill[white] svg{M86.3,186.2H70.9V79.1h15.4v48.4V186.2z}
		svg{M108.9,79.1h41.6c39.6,0,57,28.3,57,53.6c0,27.5-21.5,53.6-56.8,53.6h-41.8V79.1z M124.3,172.4h24.5c34.9,0,42.9-26.5,42.9-39.7c0-21.5-13.7-39.7-43.7-39.7h-23.7V172.4z}
		svg{M88.7,56.8c0,5.5-4.5,10.1-10.1,10.1c-5.6,0-10.1-4.6-10.1-10.1c0-5.6,4.5-10.1,10.1-10.1C84.2,46.7,88.7,51.3,88.7,56.8z};
	}
}

\newcommand\orcidicon[1]{\href{https://orcid.org/#1}{\mbox{\scalerel*{
				\begin{tikzpicture}[yscale=-1,transform shape]
				\pic{orcidlogo};
				\end{tikzpicture}
			}{|}}}}

\usepackage{graphicx,subcaption}
\graphicspath{{figures/}} 
\usepackage{transparent}
\usepackage{import}
\usepackage{algorithm,algorithmic,setspace}

\usepackage{placeins}  
\usepackage[capitalise]{cleveref}

\usepackage{relsize}
\usepackage{tikz}
\usetikzlibrary{fit,tikzmark}

\newcommand{\evec}{{\bf{e}}}

\newcommand{\yvec}{{\bf{y}}}

\newcommand{\wvec}{{\bf{w}}}
\newcommand{\xvec}{{\bf{x}}}

\newcommand{\vvec}{{\bf{v}}}
\newcommand{\gvec}{{\bf{g}}}

\newcommand{\zerovec}{{\bf{0}}}

\newcommand{\Amat}{{\bf{A}}}
\newcommand{\Bmat}{{\bf{B}}}

\newcommand{\Hmat}{{\bf{H}}}
\newcommand{\Jmat}{{\bf{J}}}
\newcommand{\Imat}{{\bf{I}}}

\def\bsigma{{\mbox{\boldmath $\Sigma$}}}

\def\thetavec{{\mbox{\boldmath $\theta$}}}

\def\thetavecsmall{{\mbox{\boldmath {\scriptsize $\theta$}}}}

\newcommand{\be}{\begin{equation}}
\newcommand{\ee}{\end{equation}}
\newcommand{\beqna}{\begin{eqnarray}}
\newcommand{\eeqna}{\end{eqnarray}}

\newcommand  \itr [2] {{{ #1}}\raisebox{1.3ex}{${\scriptscriptstyle (#2)} $}}

\newcommand \grad[1]{\nabla_{\hspace{-0.09cm}\rescalemath{0.9}{\scriptstyle#1}}} 
\newcommand \expec[1]{\mathrm{E}_{\rescalemath{0.9}{\scriptstyle#1}}} 
\newcommand \transpose[1]{#1^{\mbox{\tiny $T$}}} 			
\newcommand \rescalemath[2]{\scalebox{#1}{$\displaystyle#2$}}		
\newtheorem{theorem}{Theorem}
\newtheorem{lemma}{Lemma}

\newtheorem{proposition}{Proposition}

\def \thvec {\boldsymbol \theta}

\def \tml {\hat{ \thvec}\raisebox{1.3ex}{$\mbox{\tiny (ML)} $}}
\def \tmlx {\hat{ \thvec}_\xvec\hspace{-0.2cm}\raisebox{1.3ex}{$\mbox{\tiny (ML)} $}}
\def \tmly {\hat{ \thvec}_\yvec\hspace{-0.2cm}\raisebox{1.3ex}{$\mbox{\tiny (ML)} $}}

\newcommand \tmlm[1] {\hat{ \theta}_{#1}\hspace{-0.17cm}\raisebox{1.3ex}{$\mbox{\tiny (ML)} $}}
\def \tpsml {\hat{ \thvec}\raisebox{1.3ex}{$\mbox{\tiny (PSML)} $}}

\def \satpsml {\hat{ \thvec}\raisebox{1.3ex}{$\mbox{\tiny (SA-PSML)} $}}
\def \sbtpsml {\hat{ \thvec}\raisebox{1.3ex}{$\mbox{\tiny (2B-PSML)} $}}
\newcommand \sbtpsmlm[1] {\hat{ \theta}_{#1}\hspace{-0.17cm}\raisebox{1.3ex}{$\mbox{\tiny (2B-PSML)} $}}

\def \tjs {\hat{ \thvec}\raisebox{1.3ex}{$\mbox{\tiny (JS)} $}}
\def \tcs {\hat{ \thvec}\raisebox{1.3ex}{$\mbox{\tiny (CS)} $}}
\def \mbptpsml {\hat{ \thvec}\raisebox{1.2ex}{$\mbox{\tiny (MBP-PSML)} $}}
\def \Am {{\cal A}_m}

\crefname{subsection}{Subsection}{Subsections}
\Crefname{appendix}{Appendix}{Appendices}
\Crefname{objective}{Objective}{Objectives}
\Crefname{condition}{condition}{conditions}

\newcommand{\Deltavec}{{\bf{\Delta}}}

\newcommand\norm[1]{\left\lVert#1\right\rVert}
\title{Low-Complexity Methods for Estimation After Parameter Selection}
\author{Nadav~Harel$^\textsuperscript{\orcidicon{0000-0001-5104-2786}}$,~\IEEEmembership{Student Member,~IEEE},
	and~Tirza~Routtenberg$^\textsuperscript{\orcidicon{0000-0002-7238-7764}}$,~\IEEEmembership{Senior Member,~IEEE} 
\thanks{Nadav Harel and Tirza~Routtenberg are with the Department of Electrical and Computer Engineering, Ben-Gurion University of the Negev,
	Beer-Sheva 84105, Israel.  e-mail: nadavhar@post.bgu.ac.il,~tirzar@bgu.ac.il
	
	This research was partially supported by the 
	ISRAEL SCIENCE FOUNDATION (ISF), Grant No. 1173/16.
	Nadav Harel has been funded by the Kreitman School of Advanced Graduate
	Studies.
 }
}
\begin{document}	\bstctlcite{IEEEexample:BSTcontrol}
	\maketitle
	\begin{abstract}
		Statistical inference of multiple parameters often involves a preliminary parameter selection stage. The selection stage has an impact on subsequent estimation, for example by introducing a selection bias. The post-selection maximum likelihood (PSML) estimator is shown to reduce the selection bias and the post-selection mean-squared-error (PSMSE) compared with conventional estimators, such as the maximum likelihood (ML) estimator. 
		However, the computational complexity of the PSML is usually high due to the multi-dimensional exhaustive search for a global maximum of the post-selection log-likelihood (PSLL) function. Moreover, the PSLL involves the probability of selection that, in general, does not have an analytical form. In this paper, we develop new low-complexity post-selection estimation methods for a two-stage estimation after parameter selection architecture.
		The methods are based on implementing the iterative maximization by parts (MBP) approach, which is based on the decomposition of the PSLL function into ``easily-optimized" and complicated parts.
		  
		The proposed second-best PSML method applies the MBP-PSML algorithm with a pairwise probability of selection between the two highest-ranked parameters w.r.t. the selection rule. 
		The proposed SA-PSML method is based on using stochastic approximation (SA) and Monte Carlo integrations to obtain a non-parametric estimation of the gradient of the probability of selection and then applying the MBP-PSML algorithm on this approximation. For low-complexity performance analysis, we develop the empirical post-selection Cram$\acute{\text{e}}$r-Rao-type lower bound. 
		Simulations demonstrate that the proposed post-selection estimation methods are tractable and reduce both the bias and the PSMSE, compared with the ML estimator, while only requiring moderate computational complexity.
	\end{abstract}
\section{Introduction}

	Parameter estimation in the presence of nuisance parameters is of great interest in many signal-processing applications \cite{kay1993fundamentals,gini1996estimation,bar2018risk}. Estimation after parameter selection refers to the problem in which the choice of the {\em parameter of interest} (and, as a result, the {\em nuisance parameters}) is made by a data-based selection rule. 
	Parameter selection, as a preliminary step in estimation problems, plays an important role in modern signal processing, communication systems, and data analysis. In cognitive radio (CR) communications \cite{haykin2005cognitive,biglieri2013principles}, for example, the selection of parameters of interest may be based on the signal to noise ratio (SNR), signal energy, or transmission rate, and the parameters to be estimated could be channel gain and noise variance of the selected channel. In speech enhancement applications \cite{grimm2016phase}, a preliminary stage of reference microphone selection is conducted.
	In neuroimaging analysis \cite{vul2009puzzlingly,rosenblatt2014not_voodoo}, a subset of voxels in the brain may be selected for further analysis based on their activity pattern in functional scans.
	Another example is in power system state estimation, which is usually made only after a subset of measurements from suspicious meters is removed \cite{drayer2019detection}. In all the above-mentioned examples, there are extensions for implementation in a two-stage manner: selection in the first stage and then, estimation based on the two stages or the second stage only.
	 In all these examples, there are scenarios of a two-stage manner, where after the first-stage selection there is a second-stage of observations that can be more focused and designed specifically for the selected parameter. For example, in medical diagnosis \cite{thall1989two,dropthelosers2009,bauer2010selection}, preliminary tests may be employed to select the best treatments for a large-scale clinical trial and then, estimators are derived for the parameters of the selected treatments.
	 In dynamic programming, the well-known sequential multi-armed bandit problem \cite{vakili2013deterministic} is based on an exploration stage, which aims to select the highest-reward arm, and then, in the second stage, additional samples are taken from the selected arm to improve the reward.

	In post-selection inference, it is well known that the selection stage has an impact on subsequent estimation, by creating coupling between parameters that originally were decoupled \cite{est_after_selection}, leading to inaccurate confidence intervals, and introducing a selection bias \cite{mukhopadhyay1994multistage,putter1968estimating,cohenSackrowitz1989,whitehead1986bias,stallard2008estimation,efron2011tweedie,reid2017post}. 
	The enlightening example by Efron \cite[Fig.~1]{efron2011tweedie} demonstrates the effect of selection bias in post-selection estimation,  which is a severe problem in data-dependent selection processes. 
	By using sequential multistage schemes, the selection bias can be reduced and substantial estimation performance gain can be achieved. 
	In this paper we consider estimation after parameter selection with a two-stage data acquisition model.
	
	Estimation methods for post-parameter-selection in multi-stage models have been discussed in various works in mathematical statistics. Bias-correction methods for specific parametric models and specific estimators have been suggested in \cite{reid2017post,efron2011tweedie,whitehead1986bias,stallard2008estimation,carreras2013shrinkage,cohenSackrowitz1989,bowden2008unbiased,robertson2016accounting,robertson2018conditionally}.
	In \cite{cohenSackrowitz1989} and its extensions (see e.g. \cite{bowden2008unbiased,robertson2016accounting,robertson2018conditionally}), the Rao-Blackwell theorem has been used to develop a uniformly minimum variance conditionally unbiased estimator for two-stage estimation of the selected mean for independent Gaussian populations. However, these specific methods are based on conditioning by strict parameter ranking, which increases the variance of the estimation error.
	
	In \cite{est_after_selection} we suggested the post-selection maximum likelihood (PSML) estimator for single-stage estimation after parameter selection and developed a novel $\Psi$-{Cram{\'e}r-Rao bound (CRB)} on the post-selection mean squared error (PSMSE).
	The PSMSE, which is the mean squared error (MSE) of the selected parameter, is widely used in the mathematical statistics literature and in practical experiment design \cite{sackrowitz1984estimation,posch2005testing,bowden2008unbiased,robertson2016accounting,robertson2018conditionally,carreras2013shrinkage,reid2017post}.
	 We and others (see e.g. \cite{est_after_selection,routtenberg2014cramer,meir2017tractable,heller2017post}) showed that conditional maximum likelihood (ML) estimators, such as the PSML, are better than the ML estimator in terms of selection bias and PSMSE. 
	Moreover, we showed that if there exists an estimator which is unbiased in the Lehmann sense and achieves the $\Psi$-CRB, then it coincides with the PSML estimator for the selected parameter.
	However, the PSML estimator is based on maximization of the post-selection log-likelihood (PSLL), which usually requires a multi-dimensional exhaustive search with a computational complexity that increases with the dimension. Furthermore, for high-dimensional data and/or multi-parameter cases, an analytic representation of the PSLL involves the probability of selection, which requires high-dimensional integration \cite{bechhofer1954single}. 
	
	In this paper, we present a new model for estimation after parameter selection in a two-stage data acquisition scheme. This model is a generalization of the classical two-stage model for independent populations \cite{cohenSackrowitz1989}.
	We derive the two-stage versions of the $\Psi$-unbiasedness in the Lehmann sense \cite{lehmann2006testing} and the PSML estimator, that extend our single-stage results in \cite{est_after_selection,routtenberg2014cramer}. In the main part of this paper, we develop practical, low-complexity, estimation methods for multivariate cases where the PSML estimator is intractable. We implement the maximization by parts (MBP) algorithm, which is based on the decomposition of the likelihood function into ``easily-optimized" and complicated parts, adapted to the specific setting of post-selection estimation.
	We show the convergence of the MBP-PSML algorithm based on the ``information dominance" of the Fisher information matrix (FIM) over the information contained in the selection approach.
	Then, we use the MBP-PSML algorithm to develop two low-complexity estimation methods. The second-best PSML method uses the probability of selection between the two highest-ranked parameters in terms of the selection rule. The stochastic approximation PSML (SA-PSML) method is based on Monte Carlo approximation of the intractable gradient of the probability of selection in the PSLL maximization then plugging it into the MBP-PSML algorithm. 
	For low-complexity performance analysis, we develop the empirical post-selection FIM (PSFIM) and the empirical CRB-type lower bound. Finally, the proposed methods are examined by numerical simulations for the linear Gaussian model, Bernoulli model and for spectrum sensing in CR communication. 
		
	It should be emphasized that the considered framework is different from estimation and regression after model selection.  In our case, the observation model is assumed to be perfectly {\em{known}} and the selection approach selects the parameter of interest. In contrast, in the derivation of post-model-selection estimation methods \cite{stoica2004model} such as regression \cite{berk2013valid,lee2016exact,tibshirani2016exact,meir2017tractable,heller2017post}, as well as the associated performance bounds \cite{meir2018modelselection,meir2019cramer,sando2002cramer}, the observation model is assumed to be {\em{unknown}} and is selected from a pool of candidate models.
	Our work is also different from \cite{chaumette2005influence,joshi2009sensor,bashan2008optimal,msechu2012sensor,berberidis2016online}, where the useful {\em{data}} has been determined according to the selection rule. In contrast, in our framework, the desired {\em{parameter}} to be estimated is selected and all the data is used for statistical inference.

	The remainder of this paper is organized as follows: in \cref{background} the mathematical model for two-stage estimation after parameter selection is presented and the theoretical background is derived. In \cref{tractable_method}, we derive low-complexity post-selection estimation methods. In \cref{tractable_PSFIM} we develop a tractable, low-complexity CRB-type bound on the PSMSE. The proposed methods are evaluated via simulations in \cref{simulations}. Our conclusions appear in \cref{Conclusion}.

	In the rest of this paper, vectors are denoted by boldface lowercase letters and matrices by boldface uppercase letters.
	The notation $\mathbbm{1}_A$ denotes the indicator function of an event $A$ and
	the identity matrix is denoted by $\Imat$.
	The operator $\norm{\cdot}$ applied to a vector denotes the standard Euclidean $l_2$-norm, while applied to a matrix denotes the induced $l_2$-norm.
	The $(m,k)$th element and the $m$th column of the matrix $\Amat$ are denoted by $[\Amat]_{m,k}$ and $[\Amat]_{:,m}$, respectively.
	The notations  $\Amat\succ\Bmat$ and $\Amat\succeq\Bmat$  imply that $\Amat-\Bmat$ is a positive definite and semidefinite matrix, respectively, where $\Amat$ and $\Bmat$ are Hermitian matrices of the same size.
	The $m$th element of the gradient vector  $\grad{\thvec} c$ is given by $\frac{\partial  c}{\partial \theta_m}$,
	where $\thvec=[\theta_1,\ldots,\theta_M]^{\mbox{\tiny $T$}}$, $c$ is an arbitrary scalar function of $\thetavec$,
	$\transpose{\grad{\thvec}}  c \triangleq \transpose{(\grad{\thvec}  c)}$, and
	$\grad{\thvec}^2  c \triangleq \grad{\thvec} \transpose{\grad{\thvec}}  c$.
	The notations ${\rm{E}}_{\thetavecsmall} [\cdot]$ and ${\rm{E}}_{\thetavecsmall} [\cdot|A]$ represent the expectation and conditional expectation operators,
	parameterized by a deterministic parameter vector $\thetavec$
	and given event $A$.
	
	\section{Two-Stage Estimation After Parameter Selection} \label{background}
	In this section, we lay the groundwork for the new methods developed in this paper. 
	In \cref{two_stage_model}, we introduce the two-stage estimation after parameter selection model. Special cases of the proposed model are presented in \cref{Special_Cases}.
	In \cref{psi_Unbiasedness} we derive the PSMSE as a performance criterion and the associated $\Psi$-unbiasedness in the Lehmann sense \cite{lehmann2006testing}. In \cref{mru_psml} we present the PSML estimator for the two-stage model.							
	\subsection{Two-stage model} \label{two_stage_model}
	Let $(\Omega,{\cal{F}},P_{\thvec})$ denote a probability space, where $\Omega$ is the observation space, $\cal{F}$ is the $\sigma$-algebra, and $P_{\thvec}$ is a probability measure on $\cal{F}$ that is parameterized by a real deterministic parameter vector $\thvec=\transpose{[\theta_1,\ldots,\theta_M]} \in \mathbb{R}^M$. This probability space is assumed to be in the Hilbert space of absolutely integrable functions w.r.t. the corresponding probability measure.
	
	We consider the problem of estimating the unknown parameter vector, $\thetavec$, based on observations from $\Omega$, gathered in two stages.
	Let $\xvec \in \Omega_\xvec$ be the first-stage observation vector with the probability density functions (pdfs), $f_\xvec(\xvec;\thvec)$. A data-based selection rule, $\Psi_\xvec:\Omega_{\xvec} \rightarrow \lbrace 1,\ldots,M \rbrace$, is a deterministic function that selects a parameter of interest based on the first-stage observation vector, $\xvec$. That is, if $\Psi_\xvec(\xvec)=m$, then the estimation goal is to estimate the selected parameter, $\theta_m$.	For the sake of simplicity of notation, in the following $\Psi_\xvec(\xvec)$ is replaced by $\Psi_\xvec$. We denote by $\Pr(\Psi_\xvec=m;\thvec)$ the probability that $\theta_m$ is the selected parameter  $\forall m = 1, \ldots, M$, where it is assumed that the deterministic sets $\Am \triangleq \{ \xvec \in \Omega_\xvec : \Psi_\xvec=m   \}$ are partitions of $\Omega_{\xvec}$. Thus, $\Pr(\Psi_\xvec=m;\thvec)=\Pr(\xvec \in \Am )$. We assume a non-redundant setting in which $\Psi_\xvec$ is not a sufficient statistic for estimating $\thetavec$ based on $\xvec$,  thus, hierarchical Bayesian model \cite{lehmann2006theory} perspective will not simplify our model.
	
\begin{figure}[htb]
			\centering
			{\includegraphics[width=0.97\linewidth]{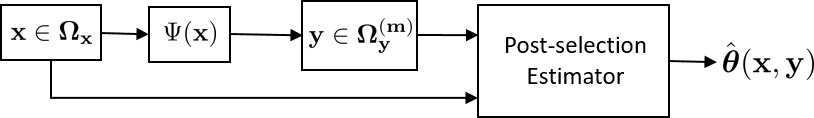}}	
		\caption{ Two-stage estimation after parameter selection scheme: First, a parameter is selected by a known predetermined selection rule, $\Psi$, based on the first-stage observation vector, $\xvec$. Then, an additional observation vector, $\yvec$, is acquired. Finally, the selected parameter is estimated based on both observation vectors, $\xvec$ and $\yvec$.}
		\label{model}
	\end{figure}

	In the second stage of data acquisition, given that the selection is $\Psi_\xvec=m$, a second observation vector, $\yvec$, is observed from $\Omega^{(m)}_{\yvec}$ with a corresponding pdf $f_m(\yvec;\thvec),~ \forall m=1,\ldots,M$. That is, we assume that the conditional pdf of $\yvec$ given $\Psi_\xvec=m$ is given by
	\be \label{pdf_y_given_x}
		f(\yvec|\Psi_\xvec=m;\thvec)=f_m(\yvec;\thvec) ~~\forall \yvec \in \Omega^{(m)}_{\yvec}.
	\ee
	It should be noted that the only influence of $\xvec$ on the second-stage observations, $\yvec$, is by choosing the generating observation model, {i.e.} the specific pdf, $f_m(\cdot)$. 
	This assumption describes a realistic scenario in which after the selection, the sample-acquisition mechanism is adapted to the selection. However, the selection by the experimenter does not change the statistical behavior of the observations, which is governed by ``nature". For example, in channel estimation, one may choose to adapt to the selection and acquire only samples from a channel that is associated with the selected parameter, but the channel's statistical behavior is the same for all the samples acquired before/after the selection.
	Therefore, by using \eqref{pdf_y_given_x} the joint pdf of the two-stage observation vectors is
	\be \label{pdf_x_y}
		f(\xvec,\yvec;\thvec)=f_{\xvec}(\xvec;\thvec)f_m(\yvec;\thvec),~~~\forall \xvec \in \Am,~ \yvec \in \Omega_\yvec,
	\ee
	where we denote $\Omega_{\yvec}\triangleq \bigcup_{m=1}^M   \Omega^{(m)}_{\yvec}$ and $\Omega \triangleq \Omega_\xvec \times \Omega_\yvec$.
	By using these definitions, \eqref{pdf_x_y}, and the rules of marginal probability, the pdf of the second-stage observation vector is given by
	 \be \label{marginal_y}  \begin{aligned}[b]	
	 	f_{\yvec}(\yvec;\thvec)=&\int\limits_{\Omega_{\xvec}}\hspace{-0.1cm}\sum_{m=1}^M  f_{\xvec}(\xvec;\thvec)f_m(\yvec;\thvec)\mathbbm{1}_{ \lbrace \xvec \in \Am \rbrace} \mathrm{d}\xvec\\
	 	=&\sum_{m=1}^{M}f_m(\yvec;\thvec)\Pr(\Psi_\xvec=m;\thvec),~ \forall \yvec \in \Omega_{\yvec}.
	 \end{aligned} 	\ee	
	Additionally, by using Bayes rule it can be verified that 
	\be \label{conditioned_likelihood_2stage}
		f(\xvec,\yvec|\Psi_\xvec=m;\thvec)=\frac{f(\xvec,\yvec;\thvec)}{\Pr(\Psi_\xvec=m;\thvec)},
	\ee
	for  $\xvec \in \Am, \yvec \in \Omega^{(m)}_{\yvec},~ \forall m=1,\ldots,M$,
	where $f(\xvec,\yvec;\thvec)$ is defined in \eqref{pdf_x_y}. 
	In addition, we define 	$f(\xvec,\yvec|\Psi_\xvec=m;\thvec)=0$, for any $\xvec\notin \Am$.
	Finally, we denote by $ \hat{ \thvec}: \Omega  \rightarrow\mathbb{R}^M$
	an estimator of $\thvec$ based on the two-stage observation vectors, $\xvec$ and $\yvec$. It should be noted that in this work we take the selection rule for granted and discuss the estimation of the selected parameter that emerged from this given selection.	The proposed two-stage estimation after parameter selection architecture is presented schematically, by in \cref{model}.

		It can be verified by using \eqref{conditioned_likelihood_2stage}, that the joint two-stage pdf from \eqref{pdf_x_y}  is essentially a finite mixture model \cite{mclachlan2004finite}:
		\be \label{pdf_x_y_new}
		f(\xvec,\yvec;\thvec)=\sum_{m=1}^M \Pr(\Psi_\xvec=m;\thvec)
		f(\xvec,\yvec|\Psi_\xvec=m;\thvec),
		\ee
		$\forall (\xvec,\yvec) \in \Omega_\xvec\times\Omega_\yvec$.
		However, in contrast to the case of a finite mixture model, where the complete likelihood function, $f(\xvec,\yvec;\thvec)$, is assumed to be or inaccessible, here it is assumed to be tractable, and may even be separable w.r.t. the unknown parameters, while the PSLL, $\log f(\xvec,\yvec|\Psi_\xvec=m;\thvec)$, which is of interest here, is the intractable part. 
	
	\subsection{Special Cases} \label{Special_Cases}
	Some special cases of the considered two-stage model from \cref{two_stage_model} are described in the following.
	\begin{enumerate}[wide, labelwidth=!, labelindent=0pt]
		\item 	{\em{Single-stage estimation after parameter selection:}} If $\Omega_{\yvec}=\varnothing$, {i.e.} there are no observations in the second stage, the two-stage model is reduced to our single-stage estimation after parameter selection model presented in \cite{est_after_selection,routtenberg2014cramer}. It should be noted that all the results in this paper are also applicable for a single-stage model. In addition, in the case where \eqref{pdf_x_y} does not hold, one can merge $\xvec$ and $\yvec$ into new single-stage vector and formulate the problem as a single-stage estimation after parameter selection.  
		\item  	 {\em{Independent populations:}} \label{independent_population}
		In many practical situations, it is common to compare several populations, select the desired one, and estimate the parameters associated with the selected population. 
		The model of two-stage estimation after selection with independent populations, which we presented in our earlier work \cite{Two_stage2016}, is a classic model in mathematical statistics (see e.g. \cite{thall1989two,stallard2008estimation,bowden2008unbiased,bauer2010selection,dropthelosers2009,cohenSackrowitz1989}). In this model, a given set of $M$ independent populations is assumed, where each population has an associated observation vector, $\xvec_m$, with a marginal pdf, $f_m(\xvec_m;\theta_m)$, parameterized by a single unknown parameter, $\theta_m$, $\forall m = 1, \ldots,M$.
		Thus, the selection of a parameter $\theta_m$ is equivalent to the selection of the $m$th population.
		 In this setting, only samples from the selected population are acquired in the second observation stage.
		 Thus, in this case, the observation pdf from \eqref{pdf_x_y} is the joint pdf of all populations:
		 \be 
		 f(\xvec,\yvec;\thvec)=\left(   \prod_{k=1}^{M} f_k(\xvec_k;\theta_k)  \right)f_m(\yvec;\theta_m), 
		 \ee
		 $~\forall \xvec \in \Am$ and $ \yvec \in \Omega_{\yvec}$.
		 In adaptive clinical trials \cite{thall1989two,stallard2008estimation,bowden2008unbiased,bauer2010selection,dropthelosers2009}, the populations may represent different medical treatments and the selection rule may select the treatment with the highest estimated life expectancy; in this case, the variance and the mean of the selected treatment are usually the parameters to estimate and the populations are usually assumed to be Gaussian. In this case, the two-stage scheme represents the different phases of medical experiments.
		 In the context of signal processing, the populations may represent independent channels for CR communications, as described in \cref{simulations_channel}, or for speech recognition \cite{wolf2014channel}. Fast multistage processing is vital for rapid wide-band sensing of channels; therefore, the two-stage model for estimation after parameter selection may provide great benefits in estimation performance.
		 	
		\item  {\em{Data-independent selection rule:}}
		The randomized selection rule satisfies $\Pr(\Psi_\xvec^{(\text{rand})}=m;\thvec)=p_m$, where $ \lbrace p_m\rbrace_{m=1}^M \in [0, 1]  $  $\forall m=1,\ldots,M$, are constant. That is, the selection of the parameter of interest is independent of the data. In particular, for $ p_m = \mathbbm{1}_{m=m_0} $, where $\theta_{m_0}$ is the parameter of interest, we obtain the well-known problem of non-Bayesian estimation in the presence of additional deterministic nuisance parameters {\cite{gini1996estimation,noammesser2009notes,bar2018risk}}.  
	\end{enumerate}
	
	\subsection{Two-stage PSMSE and $\Psi$-Unbiasedness} \label{psi_Unbiasedness}
	In this subsection, we present the PSMSE risk and its corresponding unbiasedness definition in the Lehmann sense. These definitions are an extension of similar results that we developed in \cite{est_after_selection,routtenberg2014cramer} for the single-stage model.
		
	The problem of estimation after parameter selection can be interpreted as the problem of estimation of a parameter of interest in the presence of nuisance parameters, where it is unknown in advance which is the parameter of interest; this decision is made based on the data by the selection rule. In the presence of nuisance parameters,  only estimation errors of the parameter of interest should be taken into consideration via the marginal squared-error cost of this parameter  \cite[Eq.~(1)]{bar2018risk,gini1996estimation}. Therefore, in post-selection estimation,
	the appropriate cost function is the squared error of the {\em{data-based}} selected parameter of interest, $\theta_{\Psi_\xvec(\xvec)}$:
			\be \label{SE_cost}
			\itr{C}{\Psi}( \thvec,\hat{\thvec}{(\xvec,\yvec)} )
			\triangleq (\hat{\theta}_{\Psi_\xvec(\xvec)}{(\xvec,\yvec)}-\theta_{\Psi_\xvec(\xvec)})^2 ,
			\ee
	for a given selection rule, $\Psi_\xvec$.
	By using the properties of the indicator function, the post-selection squared-error (PSSE) from \eqref{SE_cost} can be rewritten as 
	\be \label{PSSE_cost}
	\itr{C}{\Psi}\left( \thvec,\hat{\thvec}(\xvec,\yvec) \right)\triangleq \sum_{m=1}^{M}(\hat{\theta}_m(\xvec,\yvec)-\theta_m)^2 {\mathbbm{1}}_{ \lbrace \Psi_\xvec=m \rbrace}.
	\ee
	The corresponding PSMSE, which is the expected cost function, is obtained by using \eqref{PSSE_cost} and the law of total expectation:
	\be \label{psmse} \begin{aligned}[b]	 \expec{\thvec}&\hspace{-0.08cm}\left[\itr{C}{\Psi}\hspace{-0.08cm}\left( \thvec,\hat{\thvec}(\xvec,\yvec) \right)\right]=	\hspace{-0.08cm}\sum_{m=1}^{M}\hspace{-0.02cm}\expec{\thvec}\hspace{-0.1cm}\left[ (\hat{\theta}_m(\xvec,\yvec)-\theta_m)^2 {\mathbbm{1}}_{ \lbrace \Psi_\xvec=m \rbrace}   \right]\\
			=&\sum_{m=1}^{M}\expec{\thvec}\hspace{-0.08cm}\left[(\hat{\theta}_m(\xvec,\yvec)-\theta_m)^2|\Psi_\xvec=m \right]\hspace{-0.05cm}\Pr\left(\Psi_\xvec=m;\thvec\right).
	\end{aligned}
	\ee
	It can be seen that the conditional expectation on the r.h.s. of \eqref{psmse} is calculated by using the joint pdf of the two-stage observations, $f(\xvec,\yvec|\Psi_\xvec;\thvec)$ defined in \eqref{conditioned_likelihood_2stage}, while the selection probability is calculated by using $f(\xvec | \Psi_\xvec = m; \thvec)$, {i.e.} it is only a function of the pdf of the first-stage observations.
	The PSMSE in \eqref{psmse}, which is the MSE over the selected parameter, is widely used in the mathematical statistics literature and in practical experiment design \cite{sackrowitz1984estimation,posch2005testing,bowden2008unbiased,robertson2016accounting,robertson2018conditionally,carreras2013shrinkage,reid2017post}.
	
	In non-Bayesian estimation, an unbiasedness restriction is usually imposed in order to exclude trivial estimators. The Lehmann unbiasedness definition \cite{lehmann2006testing} generalizes the concept of mean-unbiasedness to unbiasedness w.r.t. the considered cost function (see e.g. \cite{routtenberg2013periodicCRB,nitzan2018cramer,meir2019cramer,bar2018risk}).
	The Lehmann unbiasedness for two-stage estimation after parameter selection w.r.t. the PSSE cost-function, named $\Psi$-unbiasedness,
	is defined as follows. 
	\begin{proposition} \label{Psi_unbiasedness} ($\Psi$-unbiasedness)
	The estimator $\hat{ \thvec}: \Omega_{\xvec}\times\Omega_{\yvec}  \rightarrow \mathbb{R}^M$ is an $\Psi$-unbiased estimator in the Lehmann sense w.r.t. the PSSE cost function if 
		\beqna \begin{aligned}[b] \label{Psi_unbiasedness_term}
			&\expec{\thvec}\left[\hat{\theta}_m(\xvec,\yvec)-\theta_m|\Psi_\xvec=m \right]=0,~~\\& \forall m=1,\ldots,M \text{ such that } \Pr\left(\Psi_\xvec=m;\thvec\right)\neq 0	.
		\end{aligned}	\eeqna
	\end{proposition}
	\begin{IEEEproof}
	This proposition can be proved along the path of the proof of Proposition 1 from \cite{est_after_selection}, by substituting the PSSE from \eqref{PSSE_cost} into the Lehmann unbiasedness definition \cite[p.~13]{lehmann2006testing}. The full proof is omitted due to space limitations.	
	\end{IEEEproof}
	It should be noted that in the $\Psi$ unbiasedness definition in \eqref{Psi_unbiasedness_term}, the estimator is a function of the two-stage data, but the conditional expectation is w.r.t. the selection event, which is only a function of the first-stage data. Thus, \Cref{Psi_unbiasedness} highlights one of the main advantages of the two-step model in the context of ``selection bias": in various single-stage models, no $\Psi$-unbiased estimator exists, while in the equivalent two-stage model there is an $\Psi$-unbiased estimator \cite{cohenSackrowitz1989,bowden2008unbiased,Two_stage2016}. In particular, we prove in the \Cref{proof_psi_unbiased_existence}that, for any setting with an existing mean-unbiased estimator without the selection, we can find an $\Psi$-unbiased estimator for the two-stage model with at least one sample at the second stage. 
	
\subsection{PSML estimator} \label{mru_psml}
	Similar to the uniformly minimum variance unbiased estimator \cite[p.~20]{kay1993fundamentals},  
	the uniformly minimum risk unbiased estimator is an estimator that is uniformly $\Psi$-unbiased and achieves minimum PSMSE, does not always exist and may be intractable.	 
	Therefore, similarly to the commonly-used ML estimator,
	\be \label{ml} 
	\tml(\xvec,\yvec)\triangleq \arg\max_{\thvec\in \mathbb{R}^M}~\log f(\xvec,\yvec;\thvec),
	\ee
	$\forall (\xvec,\yvec) \in \Omega_\xvec\times\Omega_\yvec$,
	 we define the PSML estimator as
	\be \label{psml2} 
		\tpsml(\xvec,\yvec) \triangleq  \arg\max_{\thvec\in \mathbb{R}^M}~ \log f(\xvec,\yvec|\Psi_\xvec=m;\thvec),
	\ee
	 $\forall \xvec \in \Am,~\yvec \in \Omega_\yvec$.
	 By substituting \eqref{conditioned_likelihood_2stage} in \eqref{psml2} we obtain that the PSML can be decomposed as follows:
 	\be \label{psml} \begin{aligned}[b]
 			\tpsml(&\xvec,\yvec)\\=& \arg\max_{\thvec\in \mathbb{R}^M}\hspace{0.05cm}\log f(\xvec,\yvec;\thvec)-\log \Pr(\Psi_\xvec=m;\thvec),
 	\end{aligned}
	 \ee
	  $\forall \xvec \in \Am,~\yvec \in \Omega_\yvec$. 
 	The PSML estimator from \eqref{psml} can be interpreted as a penalized ML estimator \cite{goodd1971nonparametric,eldar2004minimum}, where the penalty term is $-\log \Pr(\Psi_\xvec=m;\thvec)$, i.e. the penalty term is specifically designed to compensate for the selection approach. However, since the penalty term is not a probability density w.r.t. $\thvec$, and since we do not have any additional prior information, the PSML does not have a Bayesian interpretation.
	The maximization on the r.h.s. on \eqref{psml2} can be interpreted as the maximization step in the  expectation-maximization (EM) algorithm \cite{dempster1977EM}. However,  in contrast to EM, in the considered post-selection scheme, the likelihood is a tractable function.	
	 The PSML estimator has been shown to have better performance than the ML estimator, in terms of $\Psi$-bias and PSMSE, in various scenarios \cite{meir2017tractable,heller2017post}.
	  Moreover, it has been shown in \cite{est_after_selection}, similarly to the ML estimator and the conventional efficiency, that if an $\Psi$-efficient estimator exists, then it coincides with the PSML estimator for the selected parameter. An $\Psi$-efficient estimator, as defined in Definition 3 in \cite{est_after_selection}, is an $\Psi$-unbiased estimator that achieves the $\Psi$-CRB on the PSMSE, which is given in \cref{tractable_PSFIM}. Thus, in this case, the PSML estimator is the minimum PSMSE unbiased estimator. In addition, it was shown in \cite{andersen1970asymptotic,sen1979asymptotic} that under mild conditions the conditional ML estimator is a consistent estimator w.r.t. the conditional pdf. Thus, we can conclude that the PSML estimator from \eqref{psml} is a consistent estimator w.r.t. the conditional pdf from \eqref{conditioned_likelihood_2stage}. 
	  If the selection rule is a consistent rule, then asymptotically, the influence of the selection process on the PSML decreases, since the probability $\Pr(\Psi_\xvec=m;\thvec)$ on the r.h.s. of \eqref{psml} converges to a specific value. In this case, the PSML from \eqref{psml} coincides with the ML from \eqref{ml}. The contribution of the probability of selection, $\Pr(\Psi_\xvec=m;\thvec)$, is more significant as there is more ambiguity in the	selection process, such as in the case of close hypotheses. 
		
	We define the following regularity conditions:
	\renewcommand{\theenumi}{C.\arabic{enumi}} 
	\begin{enumerate}[wide, labelwidth=!, labelindent=0pt]
			\item \label[condition]{convexity} 
		The PSLL function, $\log f(\xvec,\yvec|\Psi_\xvec=m;\thvec)$, is a concave function w.r.t. $\thvec$. 
		\item \label[condition]{differentiability}
				The gradient vector of the PSLL function,
			$\grad{\thvec}\log f(\xvec,\yvec|\Psi_\xvec=m;\thvec)$, exists and is finite,
			$\forall \thvec \in \mathbb{R}^M$, $\xvec \in {\cal{A}}_m,~ \yvec\in\Omega_y$.
	\end{enumerate}

	Under the regularity \cref{differentiability,convexity}, the PSML estimator from \eqref{psml} can be obtained as the solution to the following score equation:
	\be \label{grad_l}
		\grad{\thvec}\log f(\xvec,\yvec;\thvec)-\gvec(\thvec)= \zerovec,~~ \forall \xvec\in \Am, \yvec\in\Omega_\yvec,
	\ee
	where the gradient of the probability of selection is defined as
	\be \label{g}
		\gvec(\thvec) \triangleq  \grad{\thvec}\log \Pr\left(\Psi_\xvec=m;\thvec\right).
	\ee
	In general, an analytical solution of \eqref{grad_l} is intractable.
	Yet, in many cases the unconditional log-likelihood function, $\log f(\xvec,\yvec;\thvec) $, is tractable, and may even be separable w.r.t. to the unknown parameters. Thus, the conventional ML can be easily found and the intractability of \eqref{grad_l} stems from:
	\renewcommand{\theenumi}{\Alph{enumi}}
	\begin{enumerate*}
		\item  the computation of $\grad{\thvec}\log \Pr(\Psi_\xvec=m;\thvec)$, which involves high-dimensional integration that does not have a closed form expression.
		\item the maximization 	may require a multi-dimensional grid search, where the computational complexity increases with the dimension of $\thetavec$.
	\end{enumerate*} 
	Hence, there is a need for practical, low-complexity estimation methods that use the special structure of the PSLL, as well as the tractability of the conventional log-likelihood part, $f(\xvec,\yvec;\thvec)$, in order to approximate the solution for the score equation from \eqref{grad_l}.
	
	\section{Low-complexity post-selection estimation methods} \label{tractable_method}
	In this section, we develop low-complexity methods for estimation after parameter selection.
	We assume that the conventional ML estimator from \eqref{ml}, which ignores the selection, is tractable and develop low-complexity methods for maximizing the PSLL. In \Cref{mbp} we apply the MBP algorithm from \cite{song2005mbp} in order to solve iteratively the optimization problem on the r.h.s. of \eqref{psml}. Since the proposed MBP-PSML algorithm requires the evaluation of the gradient of the probability of selection from \eqref{g} at any iteration point, we propose low-complexity methods that are based on the MBP algorithm: the second-best PSML method and the SA-PSML method in \cref{2ndbest} and \cref{MC_psml}, respectively.		
	\subsection{MBP-PSML} \label{mbp}
	The MBP algorithm \cite{song2005mbp} is an iterative optimization technique that divides a general log-likelihood function, $\ell(\xvec,\yvec;\thvec)$, with the observations, $\xvec$ and $\yvec$, into two parts as follows: 
	\be \label{mbp_decomposition}
		 \ell(\xvec,\yvec;\thvec)=\ell_{s}(\xvec,\yvec;\thvec)+\ell_{c}(\xvec,\yvec;\thvec),
	\ee
	where 
	 $\ell_{c}$ is the complicated, intractable part and $\ell_{s}$ is the simple part, in the sense that solving the scoring equation, $\grad{\thvec}\ell_{s}(\xvec,\yvec;\thvec)=0$, is simple. The MBP algorithm is usually initialized by the solution of this scoring equation of the simple part, $\ell_{s}$. Then, at the $i$th iteration, it evaluates the gradient of the complicated part, $\ell_{c}$, at the previous point and updates the solution by solving
	 \be \label{mbp_decomposition_step}
	\grad{\thvec}\ell_{s}(\itr{\hat{\thvec}}{i})=-\grad{\thvec}\ell_{c}(\itr{\hat{\thvec}}{i-1}).
	\ee
	 This procedure is repeated until convergence.
	Unlike other numerical methods, such as Newton-Raphson and Fisher scoring, the MBP algorithm does not require the second order derivatives of the objective function. 
		
	We apply the MBP algorithm to solve the maximization of the PSLL by using the decomposition in \eqref{psml}, where the joint log-likelihood function, is the simple part, i.e. $\ell_{s}(\xvec,\yvec;\thvec)=\log f(\xvec,\yvec;\thvec)$ and the log of the probability of selection, is set to be the complicated part, i.e. $\ell_{c}(\xvec,\yvec;\thvec)=-\log \Pr(\Psi_\xvec=m;\thvec)$. 
	Therefore, according to \eqref{mbp_decomposition_step}, the $i$th iteration of the MBP-PSML procedure updates the estimator, $\itr{\hat{\thvec}}{i}(\xvec,\yvec)$, to be the solution of
	\be \label{iteration_eq1} 
	\begin{aligned}[b]
			\grad{\thvec}\log f(\xvec,\yvec;&\itr{\hat{\thvec}}{i}(\xvec,\yvec))  
			\\=&	\grad{\thvec}\log \Pr\left(\Psi_\xvec=m;\itr{\hat{\thvec}}{i-1}(\xvec,\yvec)\right),
	\end{aligned}
	\ee
	$\forall \xvec\in \Am, \yvec\in\Omega_\yvec$.
	By substituting \eqref{g} evaluated	at $\itr{\hat{\thvec}}{i-1}(\xvec,\yvec) $ in \eqref{iteration_eq1}, the $i$th iteration estimator, $\itr{\hat{\thvec}}{i}(\xvec,\yvec)  $, can be written as the solution of:
	\be \label{iteration_eq} 
		\grad{\thvec}\log f(\xvec,\yvec;\itr{\hat{\thvec}}{i}(\xvec,\yvec))  
		=	\gvec\left(\itr{\hat{\thvec}}{i-1}(\xvec,\yvec)\right),
	\ee 
	$\forall \xvec\in \Am, \yvec\in\Omega_\yvec$.
	 The initial estimator at $i=0$ is set to be the ML estimator, $\itr{\hat{\thvec}}{0}=\tml(\xvec,\yvec)$.
	 
	 It is well known that if an efficient estimator of $\thvec$ exists, then the gradient of the log-likelihood function can be written as \cite{porat2008digital}:
	 \be \label{efficiency}
	 \grad{\thvec}\log f(\xvec,\yvec;\thvec)	=\Jmat_{\xvec,\yvec}(\thvec)\left(\tml(\xvec,\yvec)-\thvec\right),
	 \ee
	 where
		 \be \label{Jxy}
			 \Jmat_{\xvec,\yvec}(\thvec)\triangleq
			 \expec{\thvec} \left[ \grad{\thvec}\log f(\xvec,\yvec;\thvec)\transpose{\grad{\thvec}}\log f(\xvec,\yvec;\thvec)  \right]
		 \ee
	 is the conventional two-stage FIM, which is assumed to be a non-singular matrix throughout this paper.
	 By substituting the tractable term from \eqref{efficiency} in the MBP-PSML iteration from \eqref{iteration_eq} and by replacing $\Jmat_{\xvec,\yvec}^{-1}\hspace{-0.05cm}(\itr{\hat{\thvec}}{i}\hspace{-0.05cm}(\xvec,\yvec)\hspace{-0.1cm})$ with, $\Jmat_{\xvec,\yvec}^{-1}\hspace{-0.05cm}(\itr{\hat{\thvec}}{i-1}\hspace{-0.05cm}(\xvec,\yvec)\hspace{-0.1cm})$, we obtain that the MBP-PSML update iteration is linear for this special case with the following update equation:
	 \be \label{iteration_eq_efficient}
	\itr{\hat{\thvec}}{i}(\xvec,\yvec) =
	\tml(\xvec,\yvec)-\Jmat_{\xvec,\yvec}^{-1}\hspace{-0.1cm}\left(\itr{\hat{\thvec}}{i-1}\hspace{-0.05cm}(\xvec,\yvec)\hspace{-0.1cm}\right)
	\hspace{-0.05cm}\gvec\hspace{-0.1cm}\left(\itr{\hat{\thvec}}{i-1}\hspace{-0.05cm}(\xvec,\yvec)\hspace{-0.1cm}\right).
	 \ee
	 If an efficient estimator does not exist, the iteration update in \eqref{iteration_eq_efficient} can still be used as an approximation to \eqref{iteration_eq}, which is obtained by using a Taylor series, similarly to the development of the Fisher scoring method for conventional likelihood \cite[Ch.~7.7]{kay1993fundamentals}. It should be noted that under our assumption that the conventional log-likelihood is simple, the conventional FIM in \eqref{iteration_eq_efficient} is usually tractable as well, in contrast to the post-selection FIM, which is discussed in \Cref{tractable_PSFIM}. Thus, the proposed iteration in \eqref{iteration_eq_efficient} is tractable, while developing a Fisher-scoring method for the PSLL is usually intractable.
	 The MBP-PSML procedure is described in \Cref{MBP}.
	 \begin{algorithm}[htb]
	 			\setstretch{1.2}
	 	\begin{algorithmic}[1]
	 		\REQUIRE  observation vectors, $\xvec,\yvec$, convergence parameter, $\delta$.
	 		\STATE set $m=\Psi_\xvec$
	 		\STATE initialize: $i=0$, $\itr{\hat{\thvec}}{0}(\xvec,\yvec)=\tml(\xvec,\yvec)$ 
	 		\REPEAT \vspace{-0.075cm}
	 		\STATE  set $i=i+1$
	 		\STATE solve \eqref{iteration_eq} or its approximation in \eqref{iteration_eq_efficient} to obtain the next iteration: $\itr{\hat{\thvec}}{i}(\xvec,\yvec)$ 
	 		\UNTIL {  $ \norm{ \itr{\hat{\thvec}}{i}(\xvec,\yvec) -\itr{\hat{\thvec}}{i-1}(\xvec,\yvec) } \leq \delta $   } 
	 		\ENSURE MBP-PSML estimator, $\mbptpsml(\xvec,\yvec)=\itr{\hat{\thvec}}{i}(\xvec,\yvec)$	
	 	\end{algorithmic}
	 	\caption{\small MBP-PSML}
	 	\label{MBP}
	 \end{algorithm} 
 
 In the following, we establish the convergence of the MBP-PSML method, where the PSLL function is analytically known. To this end, we define additional regularity conditions:
 \renewcommand{\theenumi}{C.\arabic{enumi}} 
 \begin{enumerate}[wide, labelwidth=!, labelindent=0pt]\setcounter{enumi}{2}
 	\item \label[condition]{Hessian}
 	The Hessian matrix of the PSLL,
 	 $\grad{\thvec}^2\log f(\xvec,\yvec|\Psi_\xvec=m;\thvec) $, exists and is finite,
 	$\forall \thvec \in \mathbb{R}^M,~\xvec \in \Am,~\yvec \in \Omega_{\yvec}$.
 	\item \label[condition]{interchanged}
 	The operations of integration w.r.t. $\xvec$ and $\yvec$, and differentiation w.r.t. $\thetavec$ can be interchanged   
 	$\forall \thetavec\in\mathbb{R}^M$ for any differentiable and measurable function, $q(\xvec,\yvec,\thetavec)$:
 	\be
			\int\limits_{\Omega_\xvec} \grad{\thvec}q(\xvec,\yvec,\thetavec)\mathrm{d}\xvec
			=\grad{\thvec}\int\limits_{\Omega_\xvec}q(\xvec,\yvec,\thetavec)\mathrm{d}\xvec.
 	\ee
 \end{enumerate}
 
 		\begin{theorem}(MBP-PSML convergence)
 		Under regularity \cref{differentiability,convexity,Hessian,interchanged}, the MBP-PSML from \cref{MBP} converges to the PSML estimator from \eqref{psml}. 
 	\end{theorem}
 	\begin{IEEEproof}
 		The convergence of the MBP algorithm for the general case is discussed in \cite[Sec.~4]{song2005mbp}. It is shown that the MBP algorithm converges asymptotically to the PSML estimator under regularity \cref{differentiability,convexity,Hessian,interchanged} and under a certain ``information dominance" condition. 
 		 By substituting our two-stage estimation after parameter selection model, the information dominance condition for the MBP-PSML can be written as
 		\be \label{information_dominance}
 		\norm{\Jmat_{\xvec,\yvec}^{-1}(\thvec)	\Jmat^{(m)}_{\Psi_\xvec}(\thvec)}<1,
 		\ee 
 		where $\Jmat_{\xvec,\yvec}(\thvec)$ is defined in \eqref{Jxy} and 
 		\beqna \label{J_psi}
 		\Jmat^{(m)}_{\Psi_\xvec} \triangleq \grad{\thvec}\log \Pr\left(\Psi_\xvec=m;\thvec\right)\transpose{\grad{\thvec}}\log \Pr\left(\Psi_\xvec=m;\thvec\right),
 		\eeqna 
 		 $\forall m=1,\ldots,M$. The matrix $\Jmat^{(m)}_{\Psi_\xvec}  $ in \eqref{J_psi} can be interpreted as the Fisher information content of the selection stage.
 		We assume that the selection rule, $\Psi_\xvec$, is not a sufficient statistic for the estimation of $\thetavec$ from $\xvec$.
 		Thus, the extension of the data processing inequality for Fisher information \cite{zamir1998proof} implies that
 		\be \label{data_processing_inequality}
 		\Jmat^{(m)}_{\Psi_\xvec}  \prec	 \Jmat_{\xvec}(\thvec),
 		\ee
 		where
 		\be
 			 \label{Jx}
 			\Jmat_{\xvec}(\thvec)\triangleq
 			\expec{\thvec} \left[ \grad{\thvec}\log f(\xvec;\thvec)\transpose{\grad{\thvec}}\log f(\xvec;\thvec)  \right]
 		\ee
 		is the first-stage FIM.
 		The inequality in \eqref{data_processing_inequality} implies that the information content of the selection step, which is based on the first-stage observation vector $\xvec$, is less than the whole information contained in the first-stage observation vector, $\xvec$.
 		In addition, the extension of the data refinement inequality for Fisher information \cite{zamir1998proof} implies that the single-stage information is less than or equal to the two-stage information, {i.e.}
 		\be\label{data_refinement_inequality}
 		\Jmat_{\xvec}(\thvec) \preceq \Jmat_{\xvec,\yvec}(\thvec).
 		\ee 
 		By substituting \eqref{data_refinement_inequality} in \eqref{data_processing_inequality} we obtain 
 		\be \label{almost_information_dominance}
 		\Jmat^{(m)}_{\Psi_\xvec} \prec \Jmat_{\xvec,\yvec}(\thvec).
 		\ee
 		Since $\Jmat_{\xvec,\yvec}(\thvec)$ is assumed to be a positive definite matrix and $\Jmat^{(m)}_{\Psi_\xvec}$ is a positive semidefinite matrix, \eqref{almost_information_dominance} implies that \cite[Th.~7.7.3]{horn1990matrix}
 		\be \label{information_dominance_without_norm}
 		\Jmat_{\xvec,\yvec}^{-1}(\thvec)	\Jmat^{(m)}_{\Psi_\xvec}(\thvec)\prec \Imat,
 		\ee
 		and that \eqref{information_dominance} is satisfied, which guarantees the convergence of the MBP-PSML algorithm to the PSML estimator.
 	\end{IEEEproof}
	
	The proposed MBP-PSML algorithm requires the evaluation of $\gvec(\thvec)$ from \eqref{g} at $\thvec=\itr{\hat{\thvec}}{i}(\xvec,\yvec)$ for each iteration. Usually, high-dimensional integration is required in order to compute the probability of selection.
	In the following, we develop low-complexity methods that use the MBP-PSML algorithm but do not require the analytical form of $\gvec(\thvec)$.  
	
\subsection{Second-best PSML} \label{2ndbest}
	 In this subsection, we implement the MBP-PSML algorithm from \Cref{mbp}, where we replace the probability of selection by the probability of the selection between the two highly-ranked parameters in $\thetavec$ in terms of $\Psi_\xvec$. Thus, using the second-best scheme, the computation of the probability for the challenging $M$-parameter selection problem reduces to a much simpler task of a selection between the best two parameters.
	 For example, for the population model from \cref{Special_Cases}, if the selection rule selects the population with the largest mean, then the second-best parameter is the parameter which is associated with the second largest sample mean.
     
     For a specific observation vector, $\xvec$, let $\theta_{\tilde{m}}$ denotes the second-best parameter, i.e. the parameter that would be selected in the absence of the selected parameter. That is, $\xvec\in \tilde{\cal A}_{m,\tilde{m}}$,  where $\tilde{\cal A}_{m,\tilde{m}}$ is the subset of $\Am$ such that the second best selection is $\theta_{\tilde{m}}$. Thus, $\Am=\bigcup_{k=1,k\neq m}^M \tilde{\cal A}_{m,k}$.
     We consider a pairwise selection between $\theta_m$ and $\theta_{\tilde{m}}$ by $\Psi$, where the selection of other parameters in $\thetavec$ is {prohibited}. We suggest replacing the probability $\Pr(\Psi_\xvec=m;\thvec)$ in the PSML from \eqref{psml} by the pairwise probability 
     \be \label{pairwise_probability}
     \Pr(\tilde{\Psi}_\xvec^{\scriptscriptstyle{(m,\tilde{m})}}=m;\thvec) \triangleq \Pr(\Psi_\xvec=m| \Psi_\xvec \in \{m,\tilde{m}\}  ;\thvec). 
     \ee
	 That is, we suggest the following second-best PSML estimator:
	 \beqna \label{2bpsml} \begin{aligned}[b]
	 	 \sbtpsml(\xvec&,\yvec) \\ \triangleq  \arg&\max_{\thvec\in \mathbb{R}^M} \hspace{0.05cm} \log f(\xvec,\yvec;\thvec)-\log \Pr(\tilde{\Psi}_\xvec^{\scriptscriptstyle{(m,\tilde{m})}}=m;\thvec),
	 \end{aligned}
	 \eeqna
$\forall \xvec \in \tilde{\cal A}_{m,\tilde{m}}, ~\yvec\in \Omega_\yvec$.

Similarly to in the case of the PSML estimator from \eqref{psml}, under the regularity \cref{differentiability,convexity}, the second-best PSML estimator from \eqref{2bpsml} can be obtained as the solution to the following score equation:
\be \label{grad_l2b}
\grad{\thvec}\log f(\xvec,\yvec;\thvec)-\tilde{\gvec}(\thvec)= \zerovec,
\ee
$\forall \xvec \in \tilde{\cal A}_{m,\tilde{m}}, ~\yvec\in \Omega_\yvec$,
where the gradient of the pairwise probability of selection is defined as
\be \label{gvec_2b}
\tilde{\gvec}(\thvec) \triangleq  \grad{\thvec}\log \Pr(\tilde{\Psi}_\xvec^{\scriptscriptstyle{(m,\tilde{m})}}=m;\thvec).
\ee 
By replacing $\gvec(\cdot)$ by $\tilde{\gvec}(\cdot)$ from \eqref{gvec_2b} in the MBP-PSML update from \eqref{iteration_eq}, we obtain that the $i$th iteration step of the second-best PSML estimator is given by
\be \label{iteration_eq_2b}
	\grad{\thvec}\log f \left( \xvec,\yvec;\itr{\hat{\thvec}}{i}(\xvec,\yvec) \right)=\tilde{\gvec}\left(\itr{\hat{\thvec}}{i-1}(\xvec,\yvec)\right),
\ee
$\forall \xvec \in \tilde{\cal A}_{m,\tilde{m}}, ~\yvec\in \Omega_\yvec$.
Similarly, the approximation from \eqref{iteration_eq_efficient} can be replaced by its second-best PSML version:
 \be \label{iteration_eq_efficient_2b} \begin{aligned}[b]
 	\itr{\hat{\thvec}}{i}(\xvec,\yvec)&\\=
 	\tml&(\xvec,\yvec)-\Jmat_{\xvec,\yvec}^{-1}\hspace{-0.1cm}\left(\itr{\hat{\thvec}}{i-1}\hspace{-0.05cm}(\xvec,\yvec)\hspace{-0.1cm}\right)
 	\hspace{-0.05cm}\tilde{\gvec}\left(\itr{\hat{\thvec}}{i-1}(\xvec,\yvec)\hspace{-0.1cm}\right)\hspace{-0.075cm},
 \end{aligned}
\ee
$\forall \xvec \in \tilde{\cal A}_{m,\tilde{m}}, ~\yvec\in \Omega_\yvec$.

In many cases, although the probability of selection, $\Pr(\Psi_\xvec=m;\thvec)$, is intractable, the probability of the pairwise selection, $\Pr(\tilde{\Psi}_\xvec^{\scriptscriptstyle{(m,\tilde{m})}}=m;\thvec)$, is tractable. By using the conditional probability properties the log-probability of selection can be decomposed as follows:
\beqna \label{joint_probability_of_selection} \begin{aligned}[b]
	\log\Pr(\Psi_\xvec=m;\thvec) =&	\log \Pr( \Psi_\xvec \in \{m,\tilde{m}\}  ;\thvec)\\&+ \log\Pr(\tilde{\Psi}_\xvec^{\scriptscriptstyle{(m,\tilde{m})}}=m;\thvec).
\end{aligned}
\eeqna
Thus, the use of $\Pr(\tilde{\Psi}_\xvec^{\scriptscriptstyle{(m,\tilde{m})}}=m;\thvec)$ in \eqref{2bpsml} instead of $\Pr(\Psi_\xvec=m;\thvec)$ is equivalent to neglecting the term $\log \Pr(  \Psi_\xvec \in \{m,\tilde{m}\}  ;\thvec)$ in the PSLL maximization. Several papers analyze scenarios and conditions where this probability is, indeed, negligible {\cite{bechhofer1954single,gupta1965some}}. However, usually this probability is non-negligible and the proposed second-best PSML is an ad-hoc method. An exception is for the trivial case when the number of parameters $M=2$, the selection is pairwise, and, $\tilde{\gvec}(\thvec)$ from \eqref{gvec_2b} coincides with $ {\gvec}(\thvec)$ from \eqref{g}. For this special case, the second-best PSML estimator coincides with the PSML estimator for this special case. 

In some cases the pairwise selection probability, $\Pr(\tilde{\Psi}_{\scriptscriptstyle{m,\tilde{m}}}(\xvec)=m;\thvec)$, is only a function of $\theta_m$ and $\theta_{\tilde{m}}$. Thus, \eqref{gvec_2b} implies that for these cases $\tilde{g}_k(\thvec)=0$,  $\forall k=1,\ldots,M$, $k\neq
m,\tilde{m}$, and $\tilde{g}_k(\thvec)=\tilde{g}_k(\theta_m,\theta_{\tilde{m}})$ for $k=m$ or $k=\tilde{m}$. By substituting these results in \eqref{iteration_eq_efficient_2b} it can be seen that at the $i$th iteration the estimator of $\thetavec$ is
given by
 \be \label{iteration_eq_efficient_2b_ornot2b} \begin{aligned}[b]
 			\itr{\hat{\thvec}}{i}&(\xvec,\yvec)=
 		\tml(\xvec,\yvec) \\
 		-&\left[\Jmat_{\xvec,\yvec}^{-1}\hspace{-0.1cm}\left(\itr{\hat{\thvec}}{i-1}(\xvec,\yvec)\hspace{-0.1cm}\right) \right]_{:,m}
 		\tilde{g}_m\left(\itr{\hat{\theta}_m\hspace{-0.2cm}}{{i-1}}(\xvec,\yvec),
 		\itr{\hat{\theta}_{\tilde{m}}\hspace{-0.2cm}}{i-1}(\xvec,\yvec)\right)\\
 			-&\left[\Jmat_{\xvec,\yvec}^{-1}\hspace{-0.1cm}\left(\itr{\hat{\thvec}}{i-1}\hspace{-0.05cm}(\xvec,\yvec)\hspace{-0.1cm}\right) \right]_{:,\tilde{m}}
 		\tilde{g}_{\tilde{m}}\left(\itr{\hat{\theta}_m\hspace{-0.2cm}}{i-1}(\xvec,\yvec),
 		\itr{\hat{\theta}_{\tilde{m}}\hspace{-0.2cm}}{i-1}(\xvec,\yvec)\right).
 \end{aligned}
\ee
According to \eqref{iteration_eq_efficient_2b_ornot2b}, in the general case, even in this special case, all the parameters should be updated in order to obtain the second-best PSML. However, for the following scenarios we can update only the selected and the second best parameters, and set all the others to their ML estimators:
\renewcommand{\theenumi}{\Alph{enumi}}
\begin{enumerate}[wide, labelwidth=!, labelindent=0pt]
	\item For the independent populations model from \cref{Special_Cases}, the FIM, $\Jmat_{\xvec,\yvec}(\thvec)$, is a diagonal
	matrix. Thus, in this scenario, \eqref{iteration_eq_efficient_2b_ornot2b} implies that only the estimators of $\theta_m$ and $\theta_{\tilde{m}}$ should be updated at each iteration via \eqref{iteration_eq_efficient_2b_ornot2b} and the other estimators of the parameters are equal to their ML value,
	\be \label{ML_non_selected}
	\sbtpsmlm{k}(\xvec,\yvec)=\tmlm{k}(\xvec,\yvec), ~\forall k \in \{1,\ldots,M\} , k \neq  \{m,\tilde{m}\}.
	\ee
	 This scenario is exemplified in the simulations in \cref{simulations_channel}.
	\item For the case where the FIM, $\Jmat_{\xvec,\yvec}(\thvec)$, is a constant matrix, such as for location family of pdfs \cite{lehmann2006testing}, the $i$th
	iteration step from \eqref{iteration_eq_efficient_2b_ornot2b} is reduced to 
	\beqna \label{iteration_eq_efficient_2b_ornot2b_const} \begin{aligned}[b]
		\itr{\hat{\thvec}}{i}(\xvec,\yvec)=&\tml(\xvec,\yvec) \\
		&-\left[  \Jmat_{\xvec,\yvec}^{-1} \right]_{:,m}
		\tilde{g}_m  \left(\itr{\hat{\theta}_m\hspace{-0.25cm}}{i-1}(\xvec,\yvec),
		\itr{\hat{\theta}_{\tilde{m}}\hspace{-0.25cm}}{i-1}(\xvec,\yvec)\right)  \\
		&-\left[  \Jmat_{\xvec,\yvec}^{-1} \right]_{:,\tilde{m}}
		\tilde{g}_{\tilde{m}} \left(\itr{\hat{\theta}_m\hspace{-0.25cm}}{i-1}(\xvec,\yvec),
		\itr{\hat{\theta}_{\tilde{m}}\hspace{-0.25cm}}{i-1}(\xvec,\yvec)\right).
	\end{aligned}
	\eeqna
	In this scenario, the update of the estimators of $\itr{\hat{\theta}}{i-1}_{\hspace{-0.6cm}m}\hspace{0.3cm}(\xvec,\yvec)$ and 
	$\itr{\hat{\theta}}{i-1}_{\hspace{-0.6cm}\tilde{m}}\hspace{0.3cm}(\xvec,\yvec)$ via \eqref{iteration_eq_efficient_2b_ornot2b_const} is not a function of the
	estimators of the other parameters. Since the PSMSE risk, as defined in \eqref{psmse}, takes into account only the estimation errors of the selected parameter, there is no need to update the non-selected parameters at each iteration that do not affect the estimation of $\theta_m$. Thus, without loss of performance, we can set these estimators to their associated ML estimators , as in \eqref{ML_non_selected}, and only update $\itr{\hat{\theta}_m\hspace{-0.2cm}}{i}(\xvec,\yvec)$ and $\itr{\hat{\theta}_{\tilde{m}}\hspace{-0.2cm}}{i}(\xvec,\yvec)$ at each iteration. This scenario is exemplified in the simulations in \cref{simulations_linear_gaussian}.
	
\end{enumerate}
 The second-best PSML algorithm is summarized in \Cref{alg:2bpsml}.

\begin{algorithm}[htb]
	\setstretch{1.2}
	\begin{algorithmic}[1]
 		\REQUIRE  observation vectors, $\xvec,\yvec$, convergence parameter, $\delta$.
		\STATE set $m$ according to $\Psi_\xvec$
		\STATE set $\tilde{m}$: the index of the parameter which would have been selected in the absence of $\theta_m$ 
		\STATE initialize: $i=0$, $\itr{\hat{\thvec}}{0}(\xvec,\yvec)=\tml(\xvec,\yvec)$ 
		\REPEAT
		\STATE  set $i=i+1$
		\STATE solve \eqref{iteration_eq_2b} or its approximation in \eqref{iteration_eq_efficient_2b} to obtain  the next iteration, $\itr{\hat{\thvec}}{i}(\xvec,\yvec) $
		\UNTIL {  $ \norm{ \itr{\hat{\thvec}}{i}(\xvec,\yvec) -\itr{\hat{\thvec}}{i-1}(\xvec,\yvec) } \leq \delta $   } 
		\ENSURE second-best estimator, $\sbtpsml(\xvec,\yvec)=\itr{\hat{\thvec}}{i}(\xvec,\yvec)$.
	\end{algorithmic}
	\caption{\small{:~Second-best PSML}}
	\label{alg:2bpsml}
\end{algorithm} \vspace{-0.5cm}

	\subsection{SA-PSML} \label{MC_psml}
In this subsection, we derive the SA method \cite{kiefer1952stochastic,blum1954multidimensional,robbinsMonro1985stochastic}. In particular, by using Monte Carlo averaging, we approximate the multi-dimensional integrals needed to calculate the gradient of the probability of selection from \eqref{g}. Then, the approximated gradient is plugged into the MBP-PSML algorithm from \cref{mbp}.
To this end, we draw samples directly from the distribution of the first-stage pdf, $f_{\xvec}(\xvec;\thvec_0)$, for a given $\thvec_0$, as described, for example, in \cite[Ch.~2]{robert2013monte}. When such generation is impossible, we use Markov chain Monte Carlo (MCMC) samplers \cite[Ch.~6]{robert2013monte} to perform the data generation step. 
 
 The probability of selecting the $m$th parameter can be written as
\be \label{probability_of_selection_integral} 
		\Pr(\Psi_\xvec=m;\thvec)
	=\int\limits_{\Omega_\xvec}\mathbbm{1}_{ \lbrace \xvec \in \Am \rbrace}f_{\xvec}(\xvec;\thvec)\mathrm{d}\xvec.
\ee
	By substituting \eqref{probability_of_selection_integral} in \eqref{g} we obtain that
		\be \label{grad_prob} \begin{aligned}[b]
			\gvec(\thvec)=\frac{\grad{\thvec} \Pr(\Psi_\xvec=m;\thvec)}{\Pr(\Psi_\xvec=m;\thvec)}
			=\frac{\displaystyle\grad{\thvec}\int\limits_{\Omega_\xvec}\mathbbm{1}_{ \lbrace \xvec \in \Am \rbrace}f_{\xvec}(\xvec;\thvec)\mathrm{d}\xvec}{\Pr(\Psi_\xvec=m;\thvec)} .
		\end{aligned}
		\ee
		Under regularity \cref{interchanged}, the operations of integration w.r.t. $\xvec$ and differentiation
		w.r.t. $\thetavec$ can be interchanged such that
	\beqna \label{swich_int_der} \begin{aligned}[b]
				\grad{\thvec}\int\limits_{\Omega_\xvec}\mathbbm{1}_{ \lbrace \xvec \in \Am \rbrace}&f_{\xvec}(\xvec;\thvec)\mathrm{d}\xvec\\
			&= \int\limits_{\Omega_\xvec}\grad{\thvec} \log f_\xvec(\xvec;\thetavec)\mathbbm{1}_{ \lbrace \xvec \in \Am \rbrace}f_{\xvec}(\xvec;\thvec)\mathrm{d}\xvec \\
			&=\expec{\thvec}\left[\grad{\thvec}\log f_{\xvec}(\xvec;\thvec)\mathbbm{1}_{ \lbrace \xvec \in \Am \rbrace}\right].
		\end{aligned}
		\eeqna
		 By substituting \eqref{swich_int_der} in \eqref{grad_prob} we obtain that
		\be \label{eq68}
				\gvec(\thvec)=\frac{\expec{\thvec}\left[\grad{\thvec}\log f_{\xvec}(\xvec;\thvec)\mathbbm{1}_{ \lbrace \xvec \in \Am \rbrace}\right]}{\Pr(\Psi_\xvec=m;\thvec)} .
		\ee
		The representation in \eqref{eq68} allows us to first calculate the gradient, $\grad{\thvec}\log f_{\xvec}(\xvec;\thvec)$, which is tractable, under our assumptions, and then use Monte Carlo evaluation of the expectation in \eqref{eq68}. To this end, for any $\thvec_0\in \mathbb{R}^M$, we draw i.i.d. samples, $\{\itr{\tilde{\xvec}}{k} \}_{k=1}^K$, from $f_{\xvec}(\xvec;\thvec_0)$, and use these samples to approximate:
		\beqna \label{gvec_indirect} 
			\gvec(\thvec_0) \approx \hat{\gvec}(\thvec_0)\triangleq 
			\frac{\displaystyle\sum_{k=1}^K \grad{\thvec}\log f_{\xvec}(\itr{\tilde{\xvec}}{k};\thvec_0)\mathbbm{1}_{ \lbrace \itr{\tilde{\xvec}}{k} \in \Am \rbrace}}{\displaystyle\sum_{k=1}^K\mathbbm{1}_{ \lbrace \itr{\tilde{\xvec}}{k} \in \Am \rbrace}}.
			\eeqna	
		In order to avoid numerical errors, if the denominator in \eqref{gvec_indirect} is smaller than a predetermined threshold, we set $\hat{\gvec}(\thvec_0)=\zerovec$. 
		From the strong law of large numbers, the approximation in \eqref{gvec_indirect} converges almost surely to $\gvec(\thvec_0),~ \forall \thvec_0\in \mathbb{R}^M$.
		
		Each iteration step of the SA-PSML method, $\itr{\hat{\thvec}}{i}(\xvec,\yvec)$, is obtained by replacing $\gvec( \cdot)$ in \eqref{iteration_eq} with with $\hat{\gvec}( \cdot)$ from \eqref{gvec_indirect}, {i.e.} as the solution of
		\be \label{iteration_eq_directSA} 
		\grad{\thvec}\log f(\xvec,\yvec;\itr{\hat{\thvec}}{i}(\xvec,\yvec))  =	\hat{\gvec}\left(\itr{\hat{\thvec}}{i-1}(\xvec,\yvec)\right).
		\ee
		Similarly, the approximation in \eqref{iteration_eq_efficient} can be used with the approximated gradient:
		\be \label{iteration_eq_efficient_sa} \begin{aligned}[b]
			\itr{\hat{\thvec}}{i}&(\xvec,\yvec)\\&=
			\tml(\xvec,\yvec)-\Jmat_{\xvec,\yvec}^{-1}\hspace{-0.1cm}\left(\itr{\hat{\thvec}}{i-1}\hspace{-0.05cm}(\xvec,\yvec)\hspace{-0.1cm}\right)
			\hspace{-0.05cm}\hat{\gvec}\left(\itr{\hat{\thvec}}{i-1}(\xvec,\yvec)\hspace{-0.1cm}\right)\hspace{-0.075cm}.
		\end{aligned}
		\ee	
	The SA-PSML method is summarized in \Cref{algSA_PSML}.		
	\begin{algorithm}[htb]
		\setstretch{1.2}
		\begin{algorithmic}[1]
			\REQUIRE  observation vectors, $\xvec,\yvec$, convergence parameter, $\delta$.
			\STATE set $m$ according to $\Psi_\xvec$
			\STATE initialize: $i=0$, $\itr{\hat{\thvec}}{0}(\xvec,\yvec)=\tml(\xvec,\yvec)$ 
			\REPEAT \vspace{-0.1cm}
			\STATE generate sample vectors $\{\itr{\tilde{\xvec}}{k}\}_{k=1}^K \sim f(\xvec;\itr{\hat{\thvec}}{i-1} (\xvec,\yvec))$\vspace{-0.5cm}
			\STATE  evaluate $\hat{ \gvec}(  \itr{\hat{\thvec}}{i-1}(\xvec,\yvec))$ from \eqref{gvec_indirect}
			\STATE solve \eqref{iteration_eq_directSA} or its approximation from \eqref{iteration_eq_efficient_sa} to obtain the next iteration: $\itr{\hat{\thvec}}{i}(\xvec,\yvec)$
			\UNTIL {  $ \norm{ \itr{\hat{\thvec}}{i}(\xvec,\yvec)-\itr{\hat{\thvec}}{i-1} (\xvec,\yvec)} \leq \delta $   } 
			\ENSURE SA-PSML estimator, $\satpsml(\xvec,\yvec)=\itr{\hat{\thvec}}{i}(\xvec,\yvec)$.
		\end{algorithmic}
		\caption{\small{:~SA-PSML}}
		\label{algSA_PSML} 
	\end{algorithm} 

	A major advantage of the SA-PSML method is that it does not require the knowledge of the selection rule, but only the ability to apply it given observations.
	That is, to compute the approximations in \eqref{gvec_indirect} we can generate the data sets, and inserts the data sets into the ``black box" to obtain the selection for each data set without the need of any other knowledge. 	
	This property is useful where the mechanism of the selection rule is not clear, or is complicated. This problem arise in experimental designs where we have access to a generative model, a {black-box} that can generate multiple realizations of the selection, while the decision rule is not clear, for example, if the selection rule is based on a chemical or biological reaction \cite{vul2009puzzlingly} whose mathematical model is not clear. Another common example is where the selection rule is based on machine learning classification algorithms \cite{sricharan2012estimation,cover1967nearest,fukunaga1990introduction} or a deep neural network classifier, where the analytic representation is usually unknown or very complicated. In \cref{knn_simulation}, we demonstrate a scenario where the selection-rule is not specifically known.
	
	\section{Empirical $\Psi$-CRB} \label{tractable_PSFIM}	
		The CRB \cite{kay1993fundamentals} provides a lower bound on the mean squared error of any mean-unbiased estimator and is used as	a benchmark in non-Bayesian estimation. However, the conventional CRB does not take into account the selection process; thus, it is inappropriate for estimation after parameter selection \cite{est_after_selection,routtenberg2014cramer,chaumette2005influence}.
		The single-stage $\Psi$-CRB was developed in \cite{est_after_selection} as an alternative, and it provides a lower bound on the PSMSE of any $\Psi$-unbiased estimator. 
		In this section we derive the extension of the single-stage $\Psi$-CRB for two-stage estimation after parameter selection. Similar to the PSML estimator, this bound may be intractable. Thus, we develop new low-complexity procedure in order to evaluate this bound.  
		
	\begin{theorem} (Two-stage $\Psi$-CRB)
		Let the regularity \Cref{differentiability,Hessian,interchanged} be satisfied,
		and $\hat{\thvec}$ be an $\Psi$-unbiased estimator of $\thvec$, with a finite second moment. Then, the PSMSE is bounded by the following $\Psi$-CRB:
		\be \label{Psi_CRB} \begin{aligned}[b]
			\expec{\thvec}\hspace*{-0.1cm}\left[C^{(\Psi)}\left( \thvec,\hat{ \thvec}(\xvec,\yvec) \right)\right]&\\  \geq \sum_{m=1}^{M}\Pr(\Psi(\xvec&)=m;\thvec)\left[{\left(  \Jmat^{(m)}_{\xvec,\yvec}(\thvec)\right)}^{-1}\right]_{m,m} 
		\end{aligned},
		\ee	
		where the post-selection FIM (PSFIM) is
		\beqna \label{Jm} \begin{aligned}[b]
			\Jmat^{(m)}_{\xvec,\yvec}(\thvec)
			\triangleq& \expec{\thvec}\left[ \grad{\thvec} \log f(\xvec,\yvec;\thvec) 
			\transpose{\grad{\thvec}} \log f(\xvec,\yvec;\thvec)| \Psi_\xvec=m \right]
			\\&-\Jmat^{(m)}_{\Psi_\xvec},
		\end{aligned} \hspace{-1cm}
		\eeqna
		$\forall m=1,\ldots,M$, and $\Jmat^{(m)}_{\Psi_\xvec}$ is defined in \eqref{J_psi}.
	\end{theorem}	

	\begin{IEEEproof}
		The proof is similar to the single-stage $\Psi$-CRB proof in \cite[Th.~1]{est_after_selection}, and can be obtained by replacing the single-stage observations pdf, $f(\xvec;\thvec)$, with the two-stage pdf, $f(\xvec,\yvec;\thvec)$.
	\end{IEEEproof}\vspace{-0.5cm}
	  	The calculation of the PSFIMs from \eqref{Jm} is often intractable due to the need for calculation of the probability of selection and the conditional expectation in \eqref{Jm}. Similarly to the empirical FIM \cite{berisha2015empirical,spall2005monte} we propose a Monte Carlo approach to approximate the PSFIMs and the $\Psi-$CRB inspired by the SA-PSML methods. The proposed SA-PSFIM utilizes the PSLL structure and, as a result, the structure of the PSFIM. 
  	By substituting \eqref{pdf_x_y}, \eqref{g}, and \eqref{J_psi} in \eqref{Jm} we obtain that
  	\be   \label{Jm_decompose}  \begin{aligned}[b]
  		\Jmat^{(m)}_{\xvec,\yvec}(\thvec)=&\expec{\thvec}\left[ \grad{\thvec} \log f_\xvec(\xvec;\thvec) 
  		\transpose{\grad{\thvec}} \log f_\xvec(\xvec;\thvec)| \Psi_\xvec=m \right]\\
  		&+2\expec{\thvec}\left[ \grad{\thvec} \log f_\xvec(\xvec;\thvec) 
  		\transpose{\grad{\thvec}} \log f_\yvec(\yvec;\thvec)| \Psi_\xvec=m \right]\\
  		&+\expec{\thvec}\left[ \grad{\thvec} \log f_\yvec(\yvec;\thvec) 
  		\transpose{\grad{\thvec}} \log f_\yvec(\yvec;\thvec)| \Psi_\xvec=m \right]\\
  		&-\gvec(\thvec)\transpose{\gvec(\thvec)}.
  	\end{aligned}
  	\ee
  	Since $\xvec$ and $\yvec$ are conditionally independent given that $\Psi_\xvec=m$,  it can be verified that
  	\beqna \label{cross_cov_J} \begin{aligned}[b]
  		\expec{\thvec}&\left[ \grad{\thvec} \log f_\xvec(\xvec;\thvec)\transpose{\grad{\thvec}} \log f_\yvec(\yvec;\thvec)| \Psi_\xvec=m \right]\\
  		&=  \expec{\thvec}\left[ \grad{\thvec} \log f_\xvec(\xvec;\thvec)| \Psi_\xvec=m \right]
  		\expec{\thvec}\left[\transpose{\grad{\thvec}} \log f_m(\yvec;\thvec) \right]\\
  		&=\zerovec,
  	\end{aligned}
  	\eeqna
  	where we use regularity \cref{interchanged}, which implies that
  	\be
  		\expec{\thvec}\left[{\grad{\thvec}} \log f_m(\yvec;\thvec) \right]=\displaystyle\grad{\thvec}\hspace{-0.2cm}\int\limits_{\Omega^{(m)}_{\yvec}}\hspace{-0.25cm}f_m(\yvec;\thvec) \mathrm{d} \yvec=\zerovec.
  	\ee
  	In addition, by using the definition in \eqref{pdf_y_given_x}, we obtain
  	\beqna \label{Jy} \begin{aligned}[b]
  		\expec{\thvec}&[ \grad{\thvec} \log f_\yvec(\yvec;\thvec) \transpose{\grad{\thvec}} \log f_\yvec(\yvec;\thvec)| \Psi_\xvec=m ]\\
  		&=\expec{\thvec}[ \grad{\thvec} \log f_m(\yvec;\thvec)\transpose{\grad{\thvec}} \log f_m(\yvec;\thvec)] \triangleq \Jmat_{\yvec}(\thvec),
  	\end{aligned}
  	\eeqna
  	which is the second-stage FIM.
  	By substituting \eqref{cross_cov_J} and \eqref{Jy} in \eqref{Jm_decompose}, we obtain
  		\be    \label{Jm2} 
		\Jmat^{(m)}_{\xvec,\yvec}(\thvec)=\Jmat^{(m)}_{\xvec}(\thvec)+\Jmat_{\yvec}(\thvec),
  	\ee
  	where 
  	\beqna \label{Jm_decompose2} \begin{aligned}[b]
  			\Jmat^{(m)}_{\xvec}(\thvec) \triangleq& \expec{\thvec}[ \grad{\thvec} \log f_\xvec(\xvec;\thvec) 
  		\transpose{\grad{\thvec}} \log f_\xvec(\xvec;\thvec)| \Psi_\xvec=m ]\\&-\gvec(\thvec)\transpose{\gvec(\thvec)}
  	\end{aligned}
  	\eeqna  	
  	is the single-stage PSFIM. 
  	Under our assumptions, the second-stage FIM, $\Jmat_{\yvec}(\thvec)$, can be analytically computed. However, the conditional expectation, as well as $\gvec(\thvec)$ on the r.h.s. of \eqref{Jm_decompose}, are intractable. Similarly to the derivation of \eqref{eq68}, we can rewrite the conditional expectation by using an indicator function as follows
  	\be \label{conditiona_var} \begin{aligned}[b]
  		\expec{\thvec}[& \grad{\thvec} \log f_\xvec(\xvec;\thvec) 
  		\transpose{\grad{\thvec}} \log f_\xvec(\xvec;\thvec)| \Psi_\xvec=m ]\\
  		&= \frac{\expec{\thvec}\left[\grad{\thvec} \log f_\xvec(\xvec;\thvec) 
  			\transpose{\grad{\thvec}} \log f_\xvec(\xvec;\thvec)\mathbbm{1}_{ \lbrace \xvec \in \Am \rbrace}\right]}{\Pr(\Psi_\xvec=m;\thvec)}.
  	\end{aligned}
  	\ee 
  	 Thus, similarly to in the case of the SA-PSML from \cref{MC_psml}, we draw $K$ i.i.d. samples, $\{\itr{\tilde{\xvec}}{k}\}_{k=1}^K $, from the true pdf, $f_{\xvec}(\xvec;\thvec)$.  We use these samples to approximate
  	\beqna \label{emp_PSFIMx} \begin{aligned}[b]
  				\Jmat^{(m)}_{\xvec}&(\thvec)\approx \hat{\Jmat}^{(m)}_{\xvec}(\thvec) \\\triangleq
  			&\frac{\displaystyle\sum_{k=1}^K \grad{\thvec}\log f_{\xvec}(\itr{\tilde{\xvec}}{k};\thvec)\transpose{\grad{\thvec}\log f_{\xvec}(\itr{\tilde{\xvec}}{k};\thvec)}\mathbbm{1}_{ \lbrace \itr{\tilde{\xvec}}{k} \in \Am \rbrace}}{\displaystyle\sum_{k=1}^K\mathbbm{1}_{ \lbrace \itr{\tilde{\xvec}}{k} \in \Am \rbrace}}
  		\\&-\hat{\gvec}(\thvec)\transpose{\hat{ \gvec}(\thvec)}.
  	\end{aligned}
  	\eeqna	
  	It can  be seen that the first term on the r.h.s. of \eqref{emp_PSFIMx} approximates the conditional expectation from \eqref{conditiona_var} and $\hat{\gvec}(\thvec)$ on the second term is approximated using \eqref{gvec_indirect}. 
  Since $\Jmat_{\yvec}(\thvec)$ can be analytically computed, we approximate
  \be \label{emp_PSFIM}
  		\hat{\Jmat}^{(m)}_{\xvec,\yvec}(\thvec) \triangleq \hat{\Jmat}^{(m)}_{\xvec}(\thvec)+\Jmat_{\yvec}(\thvec).
  \ee
   
  	The SA-PSFIM algorithm is summarized in \Cref{alg:saPSFIM68}.
  	\begin{algorithm}[htb]
  				\setstretch{1.2}
  		\begin{algorithmic}[1]
  			\REQUIRE  parameter vector $\thvec$ and the selection $m$
  			\STATE 	generate sample vectors $\{ \itr{\tilde{\xvec}}{k}\}_{k=1}^K \sim f(\xvec;\thvec)$
  			\STATE 	  evaluate $\hat{ \gvec}(\thvec)$ from \eqref{gvec_indirect}
  			\STATE  evaluate $ \hat{\Jmat}_{\xvec}^{(m)}$ from \eqref{emp_PSFIMx}
  			\STATE set $\hat{\Jmat}^{(m)}_{\xvec,\yvec}(\thvec) = \hat{\Jmat}^{(m)}_{\xvec}(\thvec)+\Jmat_{\yvec}(\thvec)$
  			\vspace{0.1cm}
  			\ENSURE $\hat{\Jmat}^{(m)}_{\xvec,\yvec}(\thvec)$.
  		\end{algorithmic}
  		\caption{\small{:~SA-PSFIM}}
  		\label{alg:saPSFIM68}
  	\end{algorithm}
  
  By substituting the  empirical SA-PSFIM from \eqref{emp_PSFIM} and the probability of selection approximated as $\frac{1}{K}\sum_{k=1}^K\mathbbm{1}_{ \lbrace \itr{\tilde{\xvec}}{k} \in \Am \rbrace}$ in \eqref{Psi_CRB}, we obtain an approximation of the $\Psi-$CRB:
  \be \label{emp_Psi_CRB}
  \hat{B}(\thvec) \triangleq \sum_{m=1}^{M}\frac{1}{K}\sum_{k=1}^K\mathbbm{1}_{ \lbrace \itr{\tilde{\xvec}}{k} \in \Am \rbrace}\left[ {\left(   \hat{\Jmat}_{\xvec,\yvec}^{(m)}(\thvec)\right)}\raisebox{1.3ex}{{$\scriptstyle -1$} }    \right]_{m,m}. 
  \ee
  By taking $K>>M$ in  \eqref{emp_PSFIM}, the SA-PSFIM can be assumed to be a non-singular matrix, and \eqref{emp_Psi_CRB} is well-defined.
    	  	
	\section{Simulations} \label{simulations}
		In this section, we evaluate the performance of the following methods:
	\renewcommand{\theenumi}{\arabic{enumi}} 
	\begin{enumerate}[wide, labelwidth=!, labelindent=0pt]
		\item The second-best PSML estimator, $\sbtpsml(\xvec,\yvec)$, from \Cref{alg:2bpsml}.
		\item The SA-PSML estimator, $\satpsml(\xvec,\yvec)$ from \cref{algSA_PSML}. 
		\item The ML estimator, $\tml(\xvec,\yvec)$, from \eqref{ml}.
		\item The split-the-data estimator, which uses only the second-stage observations, $\yvec$, for the estimation. We use the following form of the ML estimator based only on $\yvec$:
		\be \label{split_ml}
			\tmly (\yvec)\triangleq \arg\max_{\thvec\in \mathbb{R}^M}~\log f(\yvec;\thvec).
		\ee
	\item The first-stage ML estimator, 
	\be \label{ml_x}
	\tmlx (\xvec )\triangleq \arg\max_{\thvec\in \mathbb{R}^M}~\log f(\xvec;\thvec).
	\ee
	\end{enumerate}
	The performance of these estimators is compared with the empirical $\Psi$-CRB from \cref{alg:saPSFIM68}.
	The conventional CRB is not presented in this section, since it does not provide a valid bound on the PSMSE and it is significantly higher than the estimators' performance.
	
	The $\Psi$-bias and PSMSE of all estimators was calculated over $50000$ Monte Carlo simulations. The maximal number of iterations of the SA-PSML methods is limited to $50$. We set the threshold for the denominator in \eqref{gvec_indirect} to $10^{-7}\frac{{N_\xvec}^2}{M}$, while the number of generated samples is $K=1000$. The code that produces the results shown in this section is available at \url{http://www.ee.bgu.ac.il/~tirzar/publications2}.   
		
	\subsection{Linear Gaussian model}  \label{simulations_linear_gaussian}
	The linear Gaussian model with dependent or independent populations is widely used in various applications. 
	In clinical research, several treatments are compared, where each treatment has an unknown treatment effect, modeled as a Gaussian distributed variable \cite{cohenSackrowitz1989,stallard2008estimation,bowden2008unbiased,robertson2016accounting,dropthelosers2009}.
	We consider the following model with \textit{correlated} Gaussian populations:
	\be \begin{aligned}[b]
			&\xvec_n=\Hmat_{\xvec}\thvec+\wvec_n,~~~ n=1,\ldots,N_{\xvec}\\
			&\yvec_n=\itr{\Hmat}{m}_{\hspace{-0.4cm}\yvec}\hspace{0.15cm}\thvec+\vvec_n,~~~ n=1,\ldots,N_{\yvec}
	\end{aligned}~,
	\ee
	where $N_{\xvec}$ and $N_{\yvec}$ are the number of samples in the first and second stages, respectively, $\Hmat_\xvec \in \mathbb{R}^{K_\xvec \times M},\itr{\Hmat}{m}_{\hspace{-0.4cm}\yvec} \in \mathbb{R}^{K_\yvec \times M}$ are assumed to be known, full-rank matrices, where  $\itr{\Hmat}{m}_{\hspace{-0.4cm}\yvec}\hspace{0.15cm}$ is determined according to the first-stage selection from the set of known matrices, $\itr{\Hmat}{1}_{\hspace{-0.3cm}\yvec}\hspace{0.15cm},\ldots,\itr{\Hmat}{M}_{\hspace{-0.4cm}\yvec}\hspace{0.15cm}$. That is, if $\Psi	(\xvec) = m$, the second-stage data, $\yvec$, is observed with the matrix $\itr{\Hmat}{m}_{\hspace{-0.4cm}\yvec}\hspace{0.15cm}$. The noise vectors, $\lbrace \wvec_n \rbrace_{n=1}^{N_{\xvec}}$ and $\lbrace \vvec_n \rbrace_{n=1}^{N_{\yvec}}$ are statistically independent series of time-independent white Gaussian noise vectors with known covariance matrices, $\bsigma_{\wvec}, \bsigma_{\vvec}$, respectively. 
	Therefore the first-stage observation vector is $\xvec=\transpose{[ \transpose{\xvec_1},\ldots,\transpose{\xvec_{N_\xvec}} ]}$ and the second-stage observation vector is $\yvec=\transpose{[ \transpose{\yvec_1},\ldots,\transpose{\yvec_{N_\yvec}} ]}$.
	A commonly-used selection rule is the following \cite{stallard2008estimation,bowden2008unbiased,cohenSackrowitz1989}:
	\be \label{sms_selection}
 		\Psi_\xvec=\arg\max_{\hspace{-0.25cm}m=1,\ldots,M} [\tmlx (\xvec)]_m,
	\ee
	where the single-stage ML estimator from \eqref{ml_x} is given by
	\be
		\tmlx(\xvec)=\left( \transpose{\Hmat_\xvec}\bsigma_{\wvec}^{-1}\Hmat_\xvec \right)^{-1}\transpose{\Hmat_\xvec}\bsigma_{\wvec}^{-1}\bar{\xvec},
	\ee
	in which $\bar{\xvec}\triangleq \frac{1}{N_{\xvec}}\sum_{n=1}^{N_{\xvec}}\xvec_n $.
	If $\Hmat_{\xvec}=\Imat$, then, the selection rule from \eqref{sms_selection} is reduced to the commonly-used selection of the largest-mean population. The probability of selection, $\Pr(\Psi_\xvec=m;\thvec)$, for the rule in \eqref{sms_selection} is intractable for $M>2$. Thus, the PSML from \eqref{psml} cannot be directly implemented and low-complexity methods are required. 	

		In this case, split-the-data estimator from \eqref{split_ml} is given by
			\be \label{ml_y_linear}
			\tmly(\yvec)=\left(\transpose{(\itr{\Hmat}{m}_{\hspace{-0.4cm}\yvec}\hspace{0.2cm})}\bsigma_{\vvec}^{-1}\itr{\Hmat}{m}_{\hspace{-0.4cm}\yvec}\hspace{0.2cm}  \right)^{-1}\transpose{(\itr{\Hmat}{m}_{\hspace{-0.4cm}\yvec}\hspace{0.2cm})}\bsigma_{\vvec}^{-1}\bar{\yvec},		~\xvec \in \Am,
			\ee
			where $\bar{\yvec}\triangleq \frac{1}{N_{\yvec}}\sum_{n=1}^{N_{\yvec}}\yvec_n $. According to the proof in the Appendix, the estimator in \eqref{ml_y_linear} is an $\Psi$-unbiased estimator of $\thvec$.
			The ML estimator based on both $\xvec$ and $\yvec$ from \eqref{ml} is given by
		\beqna
		\tml(\xvec,\yvec)=\Jmat^{-1}_{\xvec,\yvec}  \left(    
		N_\xvec\transpose{\Hmat_\xvec}\bsigma_{\wvec}^{-1}\bar{\xvec}+
		N_\yvec \transpose{(\itr{\Hmat}{m}_{\hspace{-0.4cm}\yvec}\hspace{0.2cm})}\bsigma_{\vvec}^{-1}\bar{\yvec}   \right) ,
		\eeqna
		 $\forall \xvec \in \Am,~\yvec \in \Omega_\yvec$,
		where the conventional two-stage FIM from \eqref{Jxy} is given by
		\be
		\Jmat_{\xvec,\yvec}= N_\xvec\transpose{\Hmat_\xvec}\bsigma_{\wvec}^{-1}\Hmat_\xvec+
		N_\yvec \transpose{(\itr{\Hmat}{m}_{\hspace{-0.4cm}\yvec}\hspace{0.2cm})}\bsigma_{\vvec}^{-1}\itr{\Hmat}{m}_{\hspace{-0.4cm}\yvec}\hspace{0.15cm}  .
		\ee
		
		In order to derive the second-best PSML for this scenario, we examine the probability of pairwise selection from \eqref{pairwise_probability} for the selection rule \eqref{sms_selection},
		\beqna \label{pr_2b} \begin{aligned}[b]
			\Pr(\tilde{\Psi}_\xvec^{\scriptscriptstyle{(m,\tilde{m})}}&=m;\thvec)\\&=\Pr( \tmlm{m}(\xvec) \geq \tmlm{\tilde{m}}(\xvec))=\Phi\left( \transpose{\Deltavec_{m,\tilde{m}} }\thvec  \right),
		\end{aligned}
		\eeqna
		where $\tilde{m}$ is the index of the second-best selection, $\phi(\cdot)$ and $\Phi(\cdot)$ are the standard Gaussian pdf and cumulative distribution function (cdf), respectively, and
		\be
		\Deltavec_{m,\tilde{m}} \triangleq \frac{1}{\sqrt{\transpose{(\evec_m-\evec_{\tilde{m}})}\Jmat^{-1}_{\xvec}(\evec_m-\evec_{\tilde{m}})}}(\evec_m-\evec_{\tilde{m}}),
		\ee
		where 
		\be
			\Jmat_{\xvec}=N_{\xvec} \transpose{\Hmat_\xvec}\bsigma_{\wvec}^{-1}\Hmat_{\xvec},
		\ee
		and $\evec_m$ is the $m$th column vector of the identity matrix.
		By substituting \eqref{pr_2b} in \eqref{gvec_2b}, it can be verified that
		\be \label{gvec_2b_gauss}
		\tilde{\gvec}(\thvec)=	\frac{\phi(\transpose{\Deltavec_{m,\tilde{m}}}\thvec)}{\Phi(\transpose{\Deltavec_{m,\tilde{m}}}\thvec)}(\evec_m-\evec_{\tilde{m}}).
		\ee
		For this case, $\tml(\xvec,\yvec)$ is an efficient estimator \cite[Ch.~7]{kay1993fundamentals}. Therefore, we can use \eqref{iteration_eq_efficient} at the iteration step of the SA-PSML methods.
		By using the efficiency of this case and substituting \eqref{gvec_2b_gauss} in \eqref{iteration_eq_efficient_2b}, each iteration of the second-best PSML is given by
		\be
		\itr{\hat{\thvec}}{i}(\xvec,\yvec)=\tml(\xvec,\yvec)-\Jmat^{-1}_{\xvec,\yvec}\frac{\phi\left(\transpose{\Deltavec_{m,\tilde{m}}}\itr{\hat{\thvec}}{i-1}\right)}{\Phi\left(\transpose{\Deltavec_{m,\tilde{m}}}\itr{\hat{\thvec}}{i-1}\right)}(\evec_m-\evec_{\tilde{m}}),
		\ee
		 $\forall \xvec \in \Am,~\yvec \in \Omega_\yvec$.		
		For this Gaussian model we also compare the results with the James-Stein shrinkage estimator \cite{james1961estimation,bock1975minimax}: 
		\be
			\tjs(\xvec,\yvec)=\left( 1-\frac{M-2}{\transpose{(\tml(\xvec,\yvec))}\Jmat_{\xvec,\yvec}\tml(\xvec,\yvec)}  \right)\tml(\xvec,\yvec),
		\ee
		and the extension of the Cohen-Sackrowitz (CS) estimator \cite{cohenSackrowitz1989} for correlated populations, as described in \cite[Eq.~(2.1)]{robertson2016accounting}. It was shown in \cite{cohenSackrowitz1989,robertson2016accounting} that the CS estimator satisfies an unbiasedness condition that is stricter than our $\Psi$-unbiasedness definition from \eqref{Psi_unbiasedness_term}. Thus, the CS estimator is also an $\Psi$-unbiased estimator. However, it has strict requirements \cite{robertson2016accounting} and, thus, it has poor PSMSE performance, as shown in the following simulations. These estimators are specifically designed for the linear Gaussian model and there is no solution for the general case.

\begin{figure}[htb]
	 	\centering
		\subcaptionbox{ 	\label{lin_Gauss_bias_N}}[\linewidth]
			{\vspace{-0.05cm}\includegraphics[width=6.8cm]{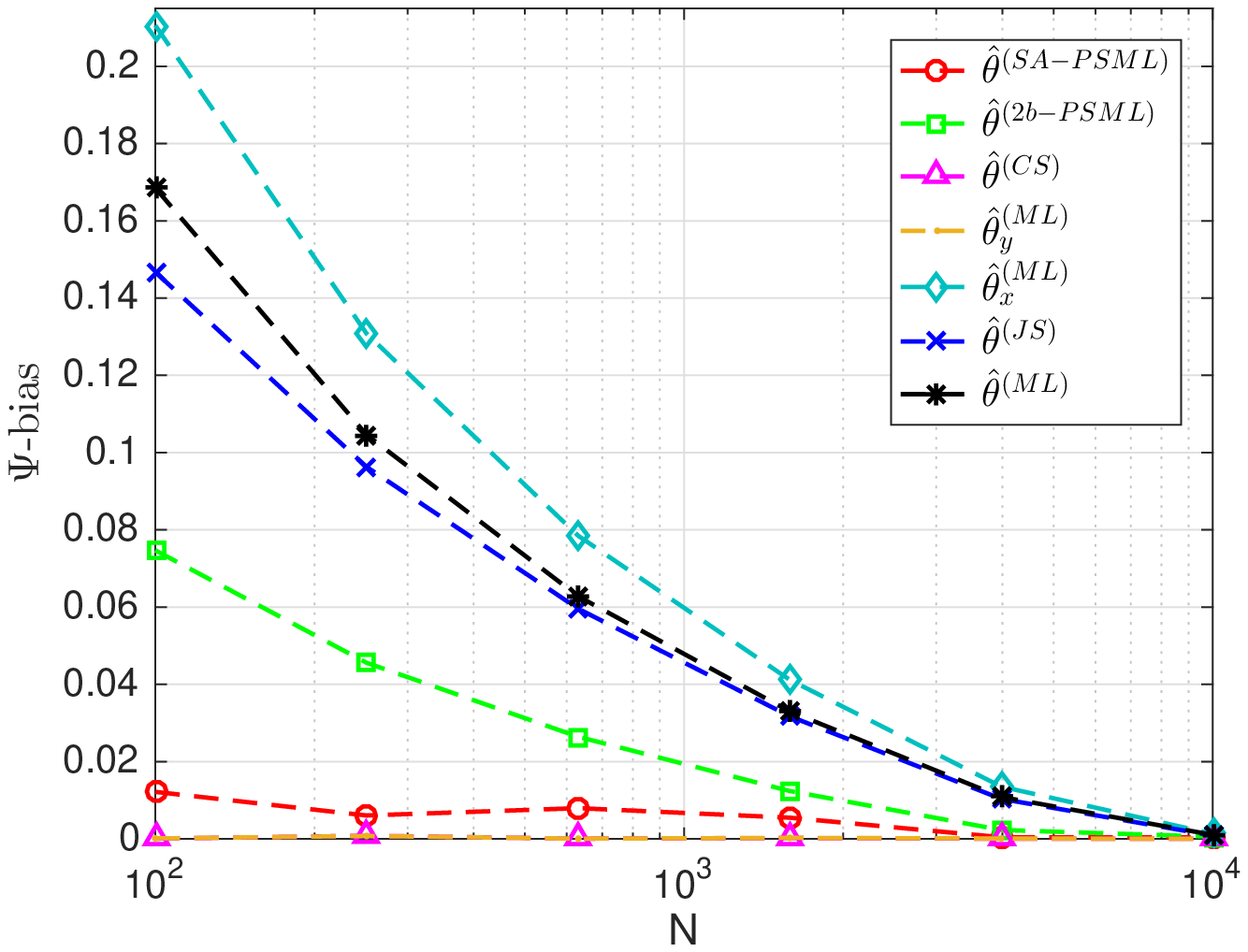}\vspace{-0.25cm}}	  
		\subcaptionbox{	\label{lin_Gauss_psmse_N}}[\linewidth]
			{\vspace{-0.05cm}\includegraphics[width=6.8cm]{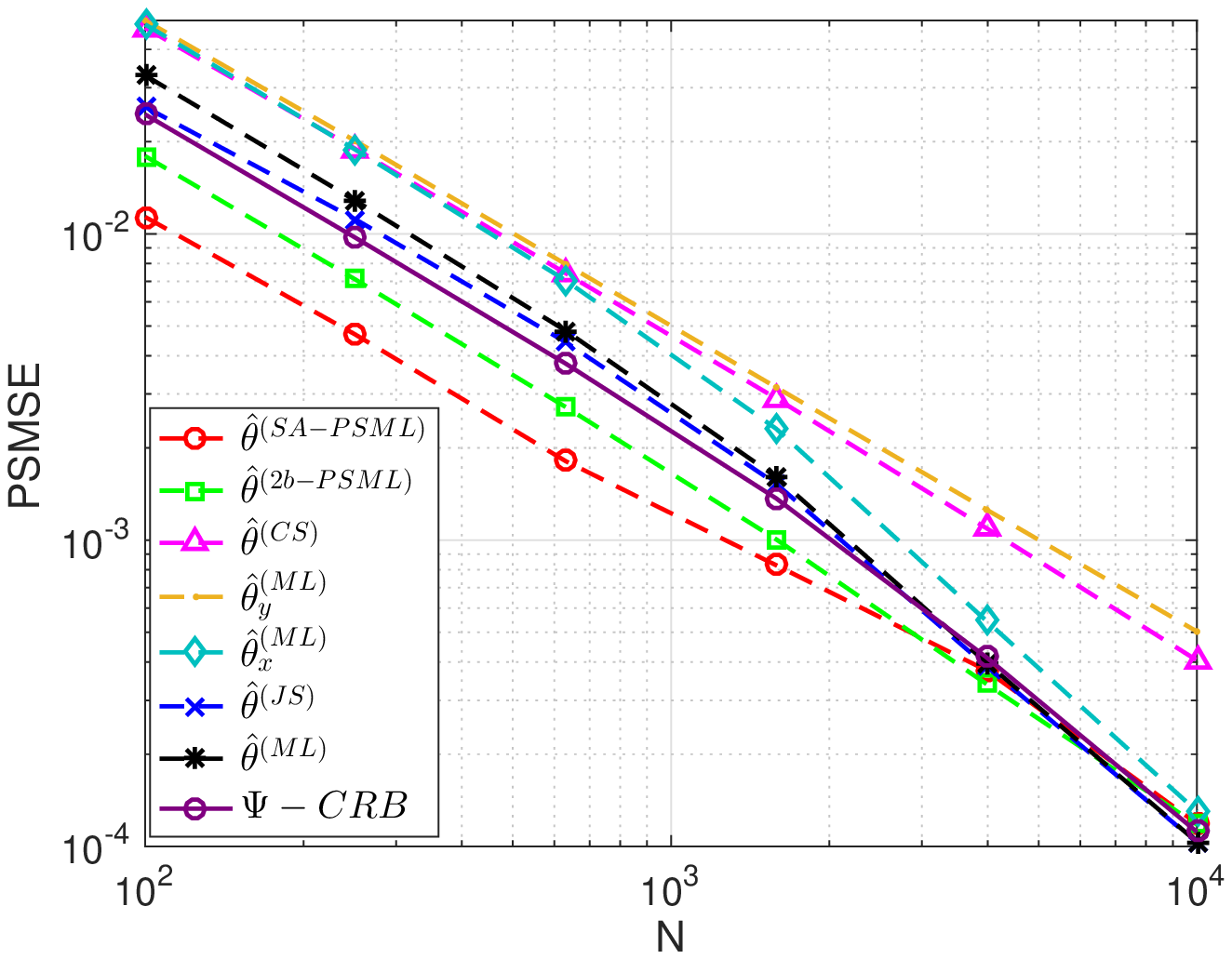}\vspace{-0.25cm}} \vspace{-0.1cm}
 	  	\caption{ Linear Gaussian model: The $\Psi$-bias (a) and PSMSE (b) of the SA-PSML, the second-best PSML, CS, split-the-data, and the ML estimator versus the number of observations, $N$.}
	 \end{figure}

		In \cref{lin_Gauss_bias_N,lin_Gauss_psmse_N} the $\Psi$-bias and PSMSE of the different estimators are presented versus the total number of observations, $N=N_\xvec+N_\yvec$, such that $N_{\xvec}=0.8N$, $N_{\yvec}=0.2N$, $M=25$, $\Hmat_\xvec=\itr{\Hmat}{m}_{\hspace{-0.4cm}\yvec}\hspace{0.15cm}=\Imat, \forall m=1,\ldots,M$, $\bsigma_{\wvec}=\bsigma_{\vvec}=\bsigma$, such that $[\bsigma]_{i,j}=(1+|i-j|)^{-2}$, and $\theta_1=1.05, \theta_2=1.01, \theta_3=1.02 ,\theta_k=1, ~k=2,\ldots,M-1, \theta_M=0$. 	
		It can be seen that the proposed PSML methods have lower $\Psi$-bias and PSMSE than the ML estimator. The CS estimator, $\tcs$, and the split-the-data estimator, $\tmly(\yvec)$, are $\Psi$-unbiased estimators, but the unbiasedness comes at the expense of the PSMSE, which is higher even than the PSMSE of the ML estimator in this case. In addition, this figure demonstrates that the empirical $\Psi$-CRB is a lower bound on the PSMSE of the $\Psi$-unbiased estimators, $\tcs$ and $\tmly(\yvec)$, and is achieved asymptotically by the PSML estimators. The James-Stein shrinkage estimator dominates the ML estimator, but it is dominated by the PSML methods. Similarly to the variance-bias trade off, the PSML methods are $\Psi$-biased, but achieve lower PSMSE than the unbiased methods and than the empirical $\Psi$-CRB. 

\begin{figure}[htb]
 	\centering
 	{\includegraphics[width=6.8cm]{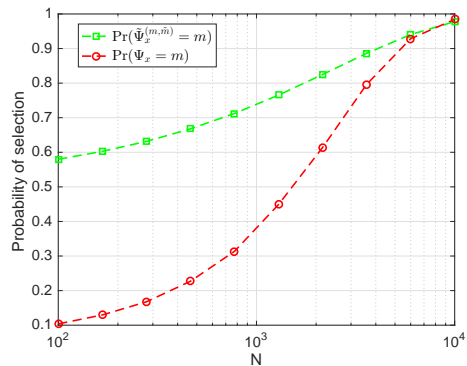}}	\vspace{-0.1cm}
 	\caption{Linear Gaussian model: Comparison between the probability of selection and the pairwise probability of selection.}
 	\vspace{-0.2cm}
 		\label{lin_Gauss_prob}		
 \end{figure}

  In \cref{lin_Gauss_prob} we compared the probability of selection, $\Pr(\Psi_\xvec=m;\thvec)$, and the pairwise probability of selection from \eqref{pairwise_probability}, $\Pr(\tilde{\Psi}_\xvec^{\scriptscriptstyle{(m,\tilde{m})}}=m;\thvec)$, for the setting of \cref{lin_Gauss_bias_N,lin_Gauss_psmse_N} for $m=1$ and $\tilde{m}=3$. The probability of selection was calculated numerically while for the pairwise probability of selection was calculated analytically according to \eqref{pr_2b}. It can be seen that these two probabilities coincide only asymptotically, which explains the advantage of the SA-PSML over the second-best PSML outside the asymptotic region. 

\begin{figure}[htb]	
		\centering
		{\includegraphics[width=6.8cm]{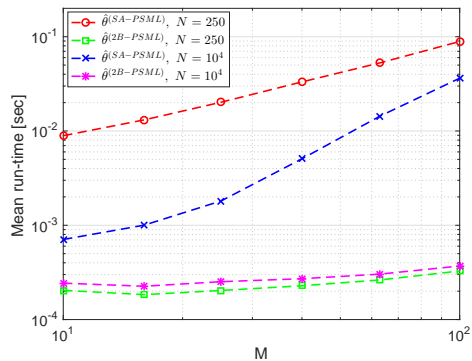}}	\vspace{-0.1cm}
		\caption{ Linear Gaussian model: Run-time of the SA-PSML and the second-best PSML methods versus the number of parameters, $M$.}
			\label{lin_Gauss_run_M}
			\vspace{-0.25cm}	
	\end{figure}

	In order to demonstrate the complexity of the proposed methods for different problem dimensions, the average processing period, ``runtime",
	is evaluated by running the algorithms using Matlab 2017b on an Intel Xeon(TM) Processor E5-2660 v4. \cref{lin_Gauss_run_M} shows the
	runtime of the PSML method versus the number of unknown parameters, $M$, for $N=250$ and $N=10^4$, and $\theta_k=1,~ \forall k=1,\ldots,M$. It can be seen that for the SA-PSML method the runtime increases with the problem dimensions.
	The second-best PSML has the lowest runtime, which is approximately constant with $M$ and with $N$ since it is based on the pairwise probability for every $M$. For larger observation number the SA-PSML requires on average fewer iterations to converge therefore the average runtime is smaller.  
	
\begin{figure}[htb]
	\centering
	\subcaptionbox{ 	\label{lin_Gauss_bias_K}}[\linewidth]
	{\vspace{-0.3cm}\includegraphics[width=6.8cm]{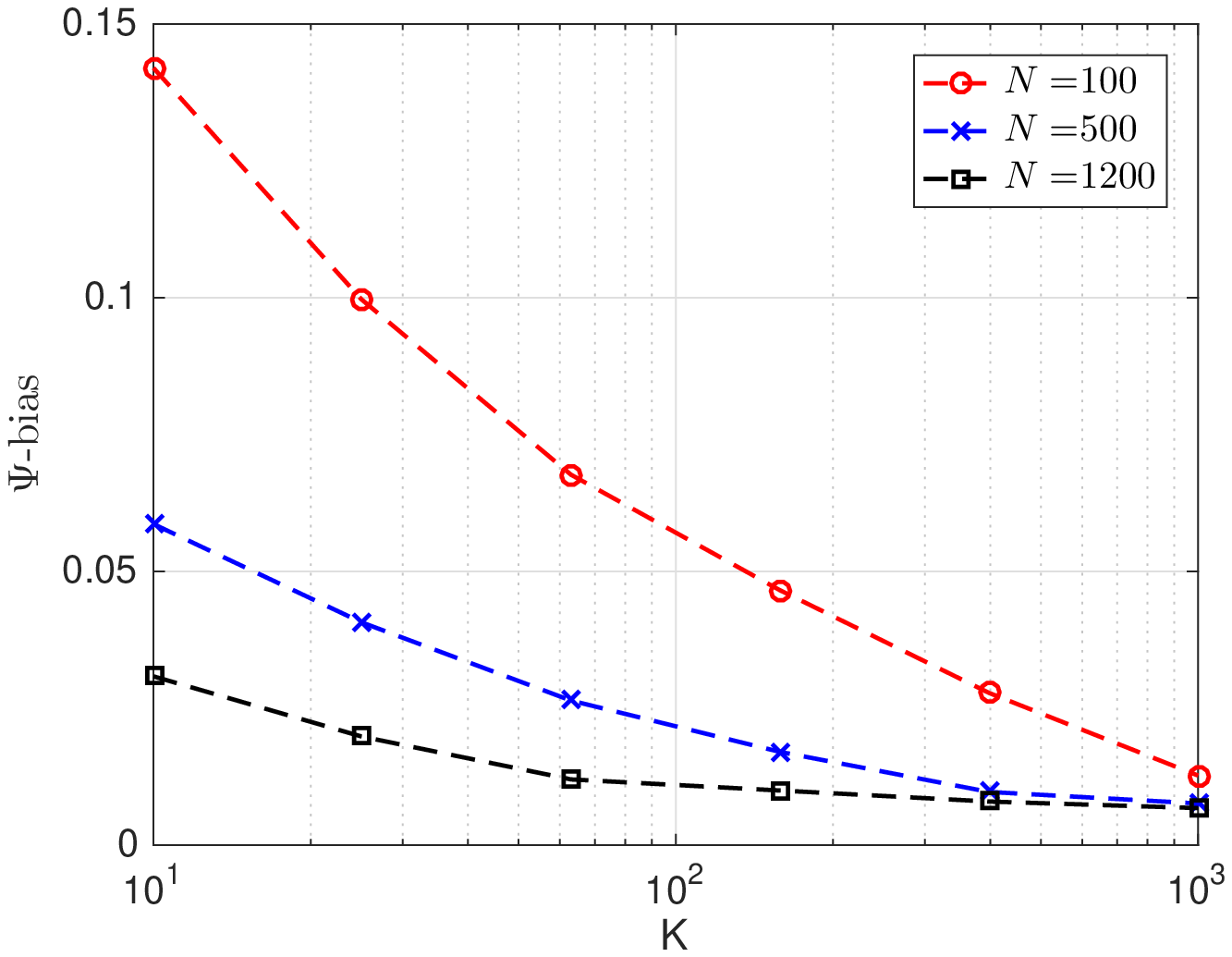}\vspace{-0.2cm}}	 
	\subcaptionbox{	\label{lin_Gauss_psmse_K}}[\linewidth]
	{\vspace{-0.05cm}\includegraphics[width=6.8cm]{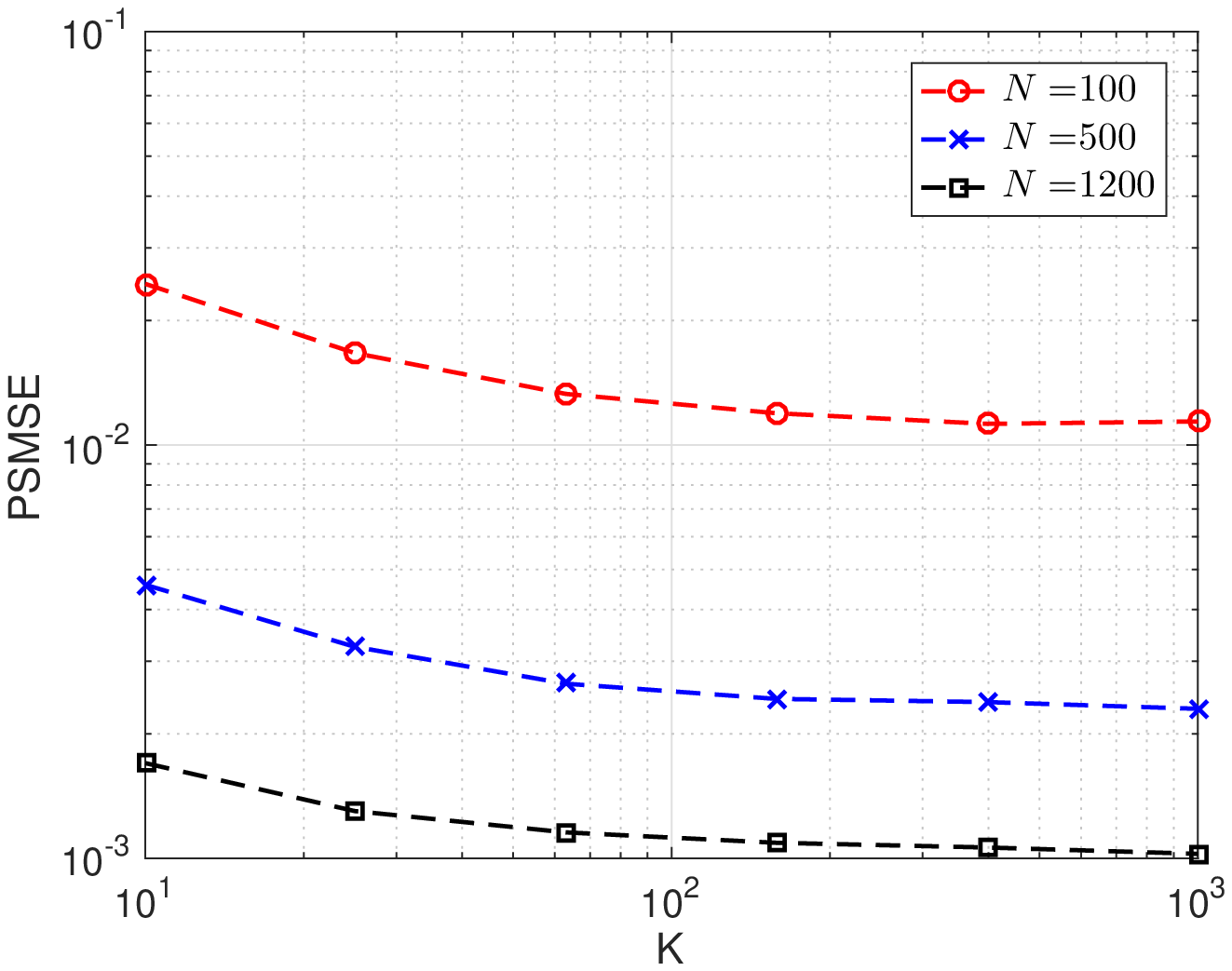}\vspace{-0.2cm}}
	\subcaptionbox{	\label{lin_Gauss_runtime_K}}[\linewidth]
	{\vspace{-0.05cm}\includegraphics[width=6.8cm]{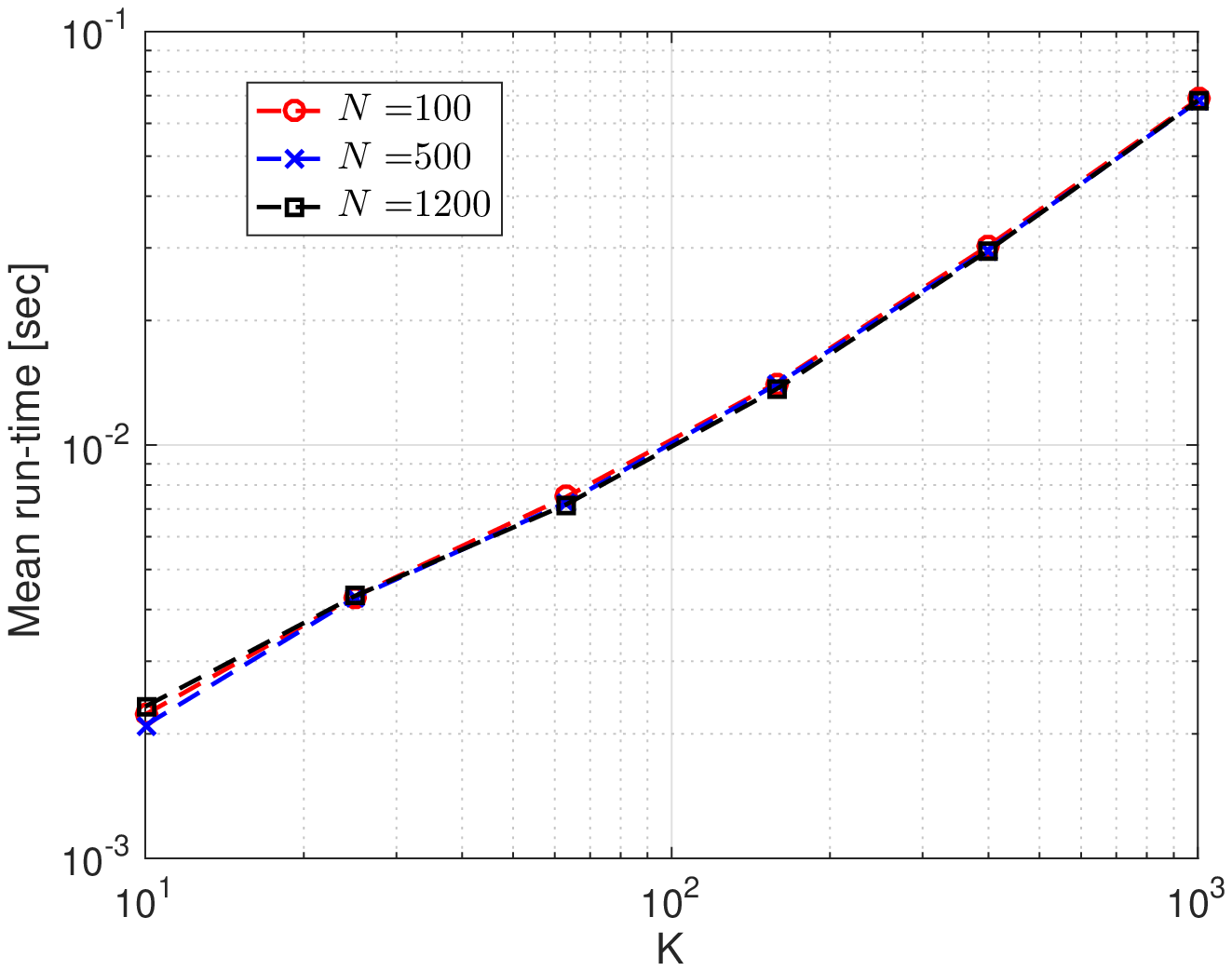}\vspace{-0.2cm}}\vspace{-0.1cm}
	\caption{Linear Gaussian model: The $\Psi$-bias (a), PSMSE (b), and runtime (c) of the SA-PSML estimator Vs. the number of Monte-Carlo simulations ,$K$.}
\end{figure}

	In \cref{lin_Gauss_bias_K,lin_Gauss_psmse_K,lin_Gauss_runtime_K} the $\Psi$-bias and PSMSE and mean runtime of the SA-PSML versus the number of Monte-Carlo simulations, $K$, for various number of observations, $N$. Although \cref{lin_Gauss_bias_K,lin_Gauss_psmse_K} exemplify the fact that as $K$ increases the SA-PSML is more accurate and the performance is better, \cref{lin_Gauss_runtime_K} shows that the computational complexity increases correspondingly. It can be seen that the influence of $N$ on the run-time is minor. 

		\subsection{Bernoulli model} \label{simulations_Bernoulli}
		We consider a Bernoulli observation model. The observation of each population yields a binary value according to a Bernoulli distribution with unknown probability of success. At the first stage all $M$ populations are observed to obtain $N_\xvec$ i.i.d. observations from every population. Based on these observations one population is selected according to a selection rule $\Psi_\xvec$. Then, another $N_\yvec$ i.i.d observations are gathered from the selected population. The goal is to estimate the probability of success of the selected population. This model is useful in multi-armed bandit problems \cite{bubeck2012regret,liu2010distributed}, where there are M arms, the $m$th arm yields a binary reward according to a Bernoulli distribution with unknown probability of success, $\theta_m$. Another example arises in medical trails \cite{dropthelosers2009}, where the response to the $m$th treatment is according to a Bernoulli distribution with unknown probability, $\theta_m$, and we would select the treatment with the highest response rate.
		Therefore, for each $n$th sample $x_k[n] \sim {{Ber}}(\theta_k),~ \forall k= 1,\ldots,M$. We denote the first-stage observation vector as $\xvec=\transpose{[x_1[1],\ldots,x_M[1],x_1[2],\ldots,x_M[2],\ldots,x_M[N_\xvec]]}$.  
		In the following, we assume that the selection rule selects the arm with the highest averaged reward, which in this case is:
		\be
		\Psi_\xvec=\arg\max_{\hspace{-0.25cm}k=1,\ldots,M} \tmlm{k}(\xvec),
		\ee 
		where the $k$th element of the single-stage ML estimator from \eqref{ml_x} is given by
		\be \label{ber_ml}
		\tmlm{k}(\xvec) \triangleq \frac{1}{N_{\xvec}}\sum_{n=1}^{N_\xvec}x_k[n],~k=1,\ldots,M,
		\ee
			$k=1,\ldots,M$.
		In the second stage, only the selected population is sampled; therefore,
		the second-stage observation vector is $\yvec=\transpose{[y_m[1],\ldots,y_m[N_\yvec]]}$, where $\xvec\in {\cal A}_m$, i.e. where the first-stage selection is $m$.
		Thus, the estimator from \eqref{split_ml}, which is the ML estimation of $\theta_{m}$ based on $\yvec$, is given by
		\be \label{berl_y}
		\tmlm{m}(\yvec) \triangleq 
			\frac{1}{N_{\yvec}}\sum_{n=1}^{N_\xvec}y_m[n],
		\ee
		and it is not defined for $k \neq m$.
		The $k$th element of the ML estimator based on both $\xvec$ and $\yvec$ is given by
		\be
		\tmlm{k}(\xvec,\yvec) \triangleq 
		\begin{cases}
			\frac{1}{N_\xvec+N_\yvec} \left( N_\xvec\tmlm{m}(\xvec)+N_\yvec{\tmlm{m}(\yvec)}  \right),\hspace*{-0.1cm} & k=m\\
			\tmlm{k}(\xvec),& k\neq m,
		\end{cases}	
		\ee
		$k=1,\ldots,M$.
		In order to derive the second-best PSML, we examine the probability of pairwise selection from \eqref{pairwise_probability}, 
		\be \label{pr_pairwise_ber}
		\Pr(\tilde{\Psi}_\xvec^{\scriptscriptstyle{(m,\tilde{m})}}=m;\thvec)
		=\Pr \left(  \tmlm{\tilde{m}}(\xvec) \hspace{0.1cm} \leq \tmlm{m}(\xvec)  \right).
		\ee 
		The random variables $N_{\xvec} \tmlm{m}(\xvec) $, $m=1,\ldots,M$ have a binomial distribution with $N_\xvec$ trials, and probability $\theta_m$. We denote the binomial probability mass function with $N$ trials and probability $\theta$ as: 
		\be \label{binomial_pmf}
		F(n; N,\theta) \triangleq  {N \choose n}  \theta^{n}(1-\theta)^{N-n}	.
		\ee
		By using \eqref{pr_pairwise_ber} and \eqref{binomial_pmf}, the probability for pairwise selection is given by
		\be \begin{aligned}[b]
			\Pr(\tilde{\Psi}_\xvec^{\scriptscriptstyle{(m,\tilde{m})}}=m;\thvec)=\sum_{n=0}^{N_\xvec} \sum_{l=0}^{n}F(n;N_\xvec,\theta_m)F(l;N_\xvec,\theta_{\tilde{m}}) 
			.
		\end{aligned}
		\ee
		One can notice that the derivative w.r.t $\theta$ of $F(n;\theta)$ is
		\be
		\frac{\partial F(n;N,\theta)}{\partial \theta}=	\xi(n,N,\theta) F(n;N,\theta),
		\ee
		where
		\be
		\xi(n,N,\theta) \triangleq \frac{n-N\theta}{\theta(1-\theta)}.
		\ee
		Therefore by substituting \eqref{pr_pairwise_ber} in \eqref{gvec_2b} we obtain that
		\be \label{gvec_2b_ber} \begin{aligned}[b]
				\tilde{\gvec}(\thvec)=
			\sum_{n=0}^{N_\xvec} \sum_{l=0}^{n}&
			\frac{F(n;N_\xvec,\theta_m)F(l;N_\xvec,\theta_k)}{\Pr(\tilde{\Psi}_\xvec^{\scriptscriptstyle{(m,k)}}=m;\thvec)} \\ \times& \left(   
			\xi(n,N_\xvec,\theta_m)\evec_m+\xi(l,N_\xvec,\theta_k)\evec_k   \right).
		\end{aligned}
		\ee 
		The two-stage FIM from \eqref{Jxy} for this scenario is a diagonal matrix, with the diagonal elements
		 $		[\Jmat_{\xvec,\yvec}(\thvec)]_{m,m}=\frac{N}{\theta_m(1-\theta_{m})}~ \forall m=1,\ldots,M$.

\begin{figure}[htb]
			\centering
			\subcaptionbox{ 	\label{binomial_bias} }[\linewidth]
			{\vspace{-0.05cm}\includegraphics[width=6.8cm]{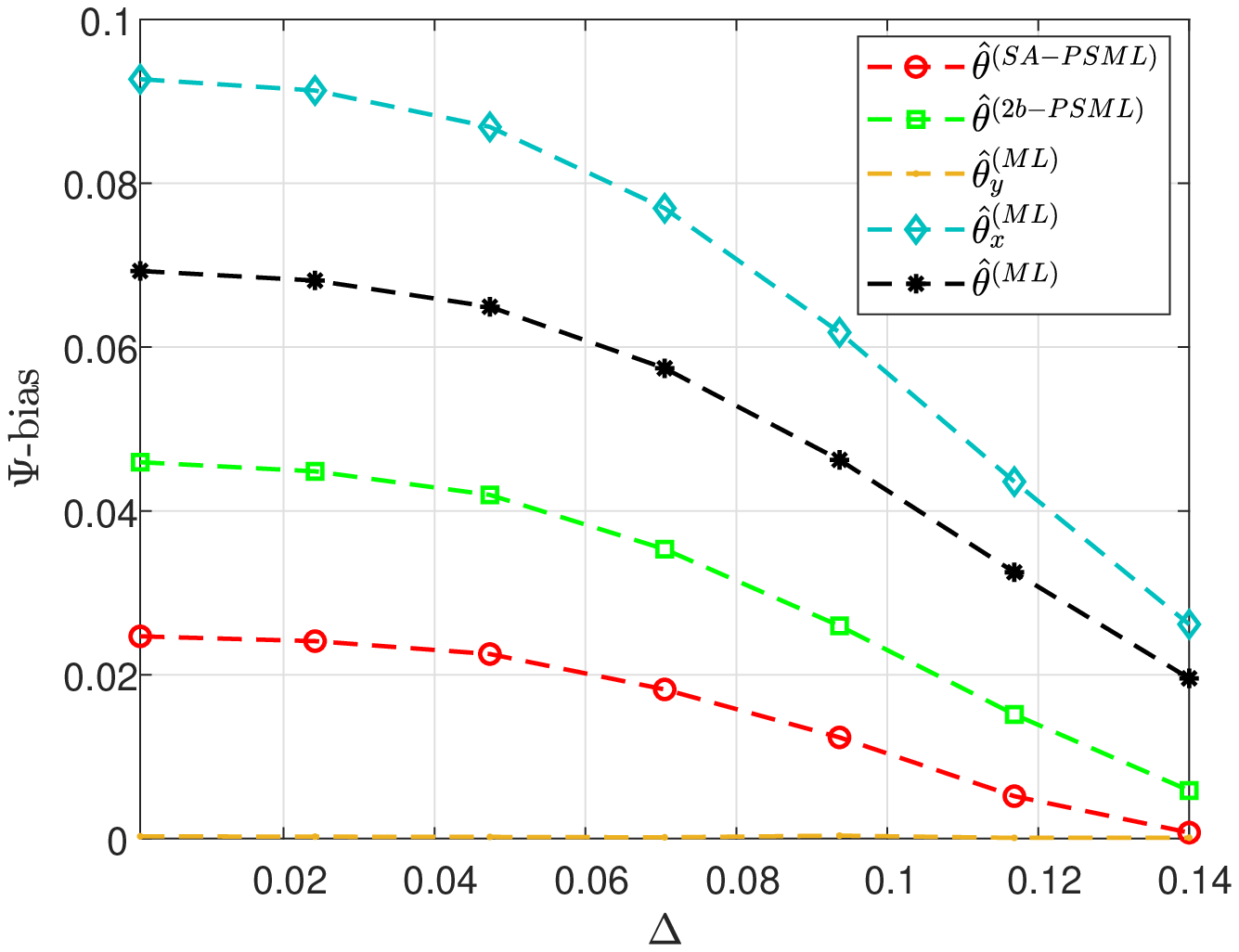}\vspace{-0.2cm}}	 
			\subcaptionbox{	\label{binomial_psmse} }[\linewidth]
			{\vspace{-0.05cm}\includegraphics[width=6.8cm]{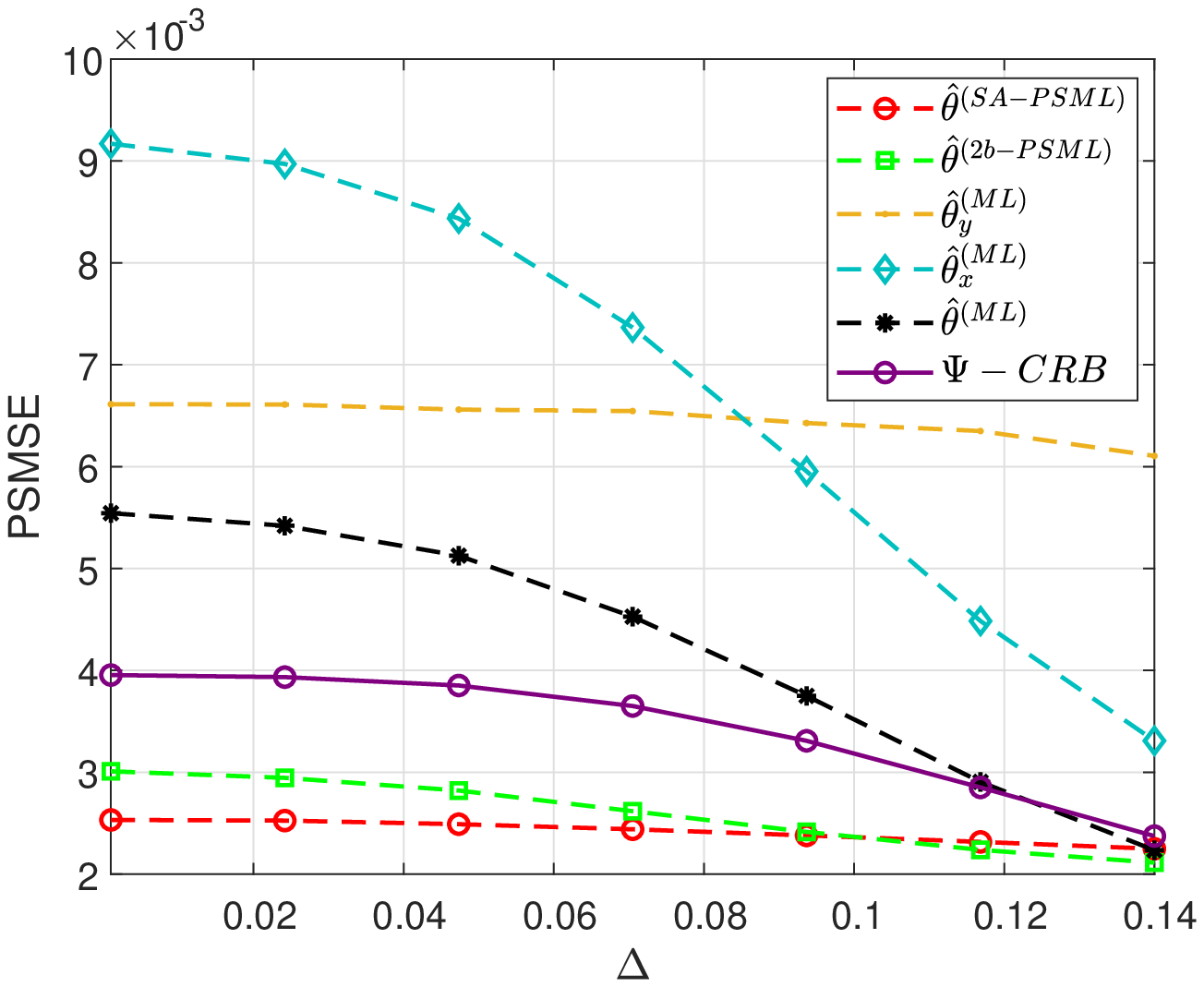}\vspace{-0.2cm}} \vspace{-0.1cm}
			\caption{ Bernoulli case: The $\Psi$-bias (a) and PSMSE (b) of the SA-PSML and second-best PSML versus $\Delta$, the difference between $\theta_1$ and the rest of the parameters, compared to split-the-data and the ML estimators.}
	\vspace{-0.5cm}	
	\end{figure}

			In \cref{binomial_bias,binomial_psmse} the $\Psi$-bias and PSMSE performance are presented versus the difference between $\theta_1$ and the rest of parameters, $\Delta$.
		In this case, $N=150$, $N_{\xvec}=0.75N$, $N_{\yvec}=0.25N$, $M=25$, and $\theta_1=0.5+\Delta,~\theta_k=0.5, ~k=2,\ldots,M$. 
		 It can be seen that the proposed PSML methods achieve better performance than the ML estimator in terms of both $\Psi$-bias and PSMSE. The split-the-data estimator, $\tml(\yvec)$, is an $\Psi$-unbiased estimator, but its PSMSE performance is the highest since it uses only part of the observations.

		\subsection{Spectrum estimation after channel selection} \label{simulations_channel}
	In this subsection, we consider a problem of two-stage spectrum estimation after channel selection. We assume a multi-channel cognitive medium access control problem \cite{haykin2005cognitive,biglieri2013principles}, where a secondary user (SU) avoids channels that are occupied by a primary user (PU). The SU should not only detect a free channel, but also choose the optimal one \cite{jiang_poor2009CR_optimal_selection}. 
	 Then, the second-stage observations set is acquired, and the goal is to estimate the parameter of the selected channel based on the two-stage observations. Unlike other studies on spectrum estimation, in which the main goal is optimal channel selection, here, we focus on the consequence estimation of the selected channel by taking into account the selection process.
	
	We assume a frequency-flat and fast-fading channel; therefore, the discrete-time input-output relation of the $k$th channel $k= 1,\ldots,M$ is given by
	\be \begin{aligned}[b]
		&x_k[n]=h_ks^{(1)}_k[n]+w^{(1)}_k[n],~~~ n=1,\ldots,N_{\xvec}\\
		&y_k[n]=h_ks^{(2)}_k[n]+w^{(2)}_k[n],~~~ n=1,\ldots,N_{\yvec}
	\end{aligned}~,
	\ee
	where $ h_k \in \lbrace 0,1 \rbrace,~\forall k= 1,\ldots,M$ are unknown deterministic parameters that represent the state of the channels. That is, $h_k=1$ indicates that the $k$th channel is occupied by a PU and $h_k=0$ indicates that the $k$th channel is free for transmission. The state parameters, $h_k,~k= 1,\ldots,M$ are considered to be constant over the sensing period. The signals $s^{(i)}_k[n],~ i={1,2}$, and the additive noise, $w^{(i)}_k[n],~ i={1,2}$, are mutually independent i.i.d. Gaussian signals, with zero mean and unknown variances, $\sigma_{s_k}^2$ and $\sigma_{w_k}^2$, respectively. Therefore, $x_k[\cdot], y_k[\cdot]\sim {\cal{N}}(0,\sigma_k^2),~ \forall k= 1,\ldots,M$, where $\sigma_k^2 \triangleq h_m\sigma_{s_k}^2+\sigma_{w_k}^2 $ and $\thvec \triangleq  \transpose{[\sigma_1^2,\ldots,\sigma_M^2]}$ is the unknown parameter vector that characterizes the channels. We denote the first-stage observation vector as $\xvec=\transpose{[x_1[1],x_2[1],\ldots,x_M[1],\ldots,x_M[N_\xvec]]}$. 
	
	A widely applied spectrum sensing technique in CR is the minimum energy selection rule \cite{biglieri2013principles,urkowitz1967energy,quan_poor2009optimal},
	\be
	\Psi_\xvec=\arg\min_{\hspace{-0.25cm}k=1,\ldots,M} \tmlm{k}(\xvec),
	\ee 
	where $k$th element of the single-stage ML estimator from \eqref{ml_x} is given by
	\be \label{sig_ml}
	\tmlm{k}(\xvec) \triangleq \frac{1}{N_{\xvec}}\sum_{n=1}^{N_\xvec}x_k^2[n],~k=1,\ldots,M,
	\ee
	$k=1,\ldots,M$.
	In this scenario we assume that in the second stage, only observations from the selected channel are taken; therefore,
	the second-stage observation vector is $\yvec=\transpose{[y_m[1],\ldots,y_m[N_\yvec]]}$, where $\xvec\in {\cal A}_m$, i.e. where the first-stage selection is $m$.
	Thus, the estimator from \eqref{split_ml}, which is the ML estimation of $\theta_{m}$ based on $\yvec$, is given by
	\be \label{sig_ml_y}
	{\tmlm{m}(\yvec)} = 
		\frac{1}{N_{\yvec}}\sum_{n=1}^{N_\xvec}y_m^2[n], 
	\ee
	and $k$th element of the ML estimator based on both $\xvec$ and $\yvec$ from \eqref{ml} is given by
	\be
		{\tmlm{k}(\xvec,\yvec)} \triangleq 
		\begin{cases}
			\frac{1}{N_\xvec+N_\yvec}\left( N_\xvec{\tmlm{m}(\xvec)}+N_\yvec{\tmlm{m}(\yvec)}\right), \hspace*{-0.1cm}&k=m\\
			\tmlm{k}(\xvec),&k\neq m.
		\end{cases}	
	\ee
	In order to derive the second-best PSML, we examine the probability of pairwise selection from \eqref{pairwise_probability}, 
	\be \label{pr_snr}
	\Pr(\tilde{\Psi}_\xvec^{\scriptscriptstyle{(m,\tilde{m})}}=m;\thvec)
	=\Pr \left(  \frac{\tmlm{m}(\xvec)}{\tmlm{\tilde{m}}(\xvec)} \leq 1  \right),
	\ee 
	where $\tilde{m}$ is the index of the second-best parameter.
	The random variables $\frac{N_{\xvec}}{\sigma^2_m} \tmlm{m}(\xvec) $, $m=1,\ldots,M$ have a central $\chi$-square distribution with $N_\xvec$ degrees of freedom, and thus, $ \frac{{\sigma^2_k}\tmlm{m}(\xvec)}{{\sigma^2_m}\tmlm{\tilde{m}}(\xvec)} $ have a $F$-central distribution \cite[Ch.~2]{kay1998fundamentals}. Therefore, by using \eqref{pr_snr}, the probability for pairwise selection of the first selection over the second is given by
	\be
	\Pr(\tilde{\Psi}_\xvec^{\scriptscriptstyle{(m,\tilde{m})}}=m;\thvec)= F \left( \frac{\sigma^2_{\tilde{m}}}{\sigma^2_m}  \right),
	\ee
	and the derivative of its log w.r.t. $\thvec$ ,from \eqref{gvec_2b} is
	\be \label{gvec_2b_CE}
	\tilde{\gvec}(\thvec)=
	\frac{\varphi(\zeta)}{\theta_mF(\zeta)}(\evec_{\tilde{m}}-\zeta\evec_m),
	\ee
	where $F(\cdot)$ and $\varphi(\cdot)$ are the standard cdf and pdf of the $F$ distribution, respectively, and $\zeta \triangleq \frac{\theta_{\tilde{m}}}{\theta_m}$. 
	The two-stage FIM from \eqref{Jxy} for this scenario is a diagonal matrix, where its diagonal elements are given by $
	[\Jmat_{\xvec,\yvec}(\thvec)]_{m,m}=\frac{N}{2\theta_m^{2}} \forall m=1,\ldots,M$.
	Since $\tml(\xvec,\yvec)$ is an efficient estimator, by substituting \eqref{gvec_2b_CE} in \eqref{iteration_eq_efficient_2b} the iteration of the second-best PSML using MBP-PSML is obtained by
	\beqna \begin{aligned}[b]
		\itr{\hat{\thvec}}{i}&(\xvec,\yvec) = \tml(\xvec,\yvec)\\ 
		&-\Jmat^{-1}_{\xvec,\yvec}\hspace{-0.05cm}{\left(\hspace{-0.05cm}\itr{\hat{\thvec}}{i-1}(\xvec,\yvec)\hspace{-0.1cm}\right)}
		\rescalemath{0.9}{\frac{\varphi(\rescalemath{0.9}{\itr{\zeta}{i-1}})}{\itr{\hat{\theta}}{i-1}_{\hspace{-0.6cm}m}\hspace{0.3cm}F(\rescalemath{0.9}{\itr{\zeta}{i-1}})}}
		\hspace{-0.05cm}(\evec_{\tilde{m}}-\rescalemath{0.9}{\itr{\zeta}{i-1}}\evec_m),
	\end{aligned}
	\eeqna
	$\forall \xvec \in \Am,~\yvec \in \Omega_\yvec$,
	where $\itr{\zeta}{i} \triangleq \frac{\itr{\hat{\theta}}{i}_{\hspace{-0.3cm}\tilde{m}}}{\itr{\hat{\theta}}{i}_{\hspace{-0.3cm}m}}~$.
	
\begin{figure}[htb]
		\centering
		\subcaptionbox{ 	\label{CE_bias_N} }[\linewidth]
		{\vspace{-0.05cm}\includegraphics[width=6.8cm]{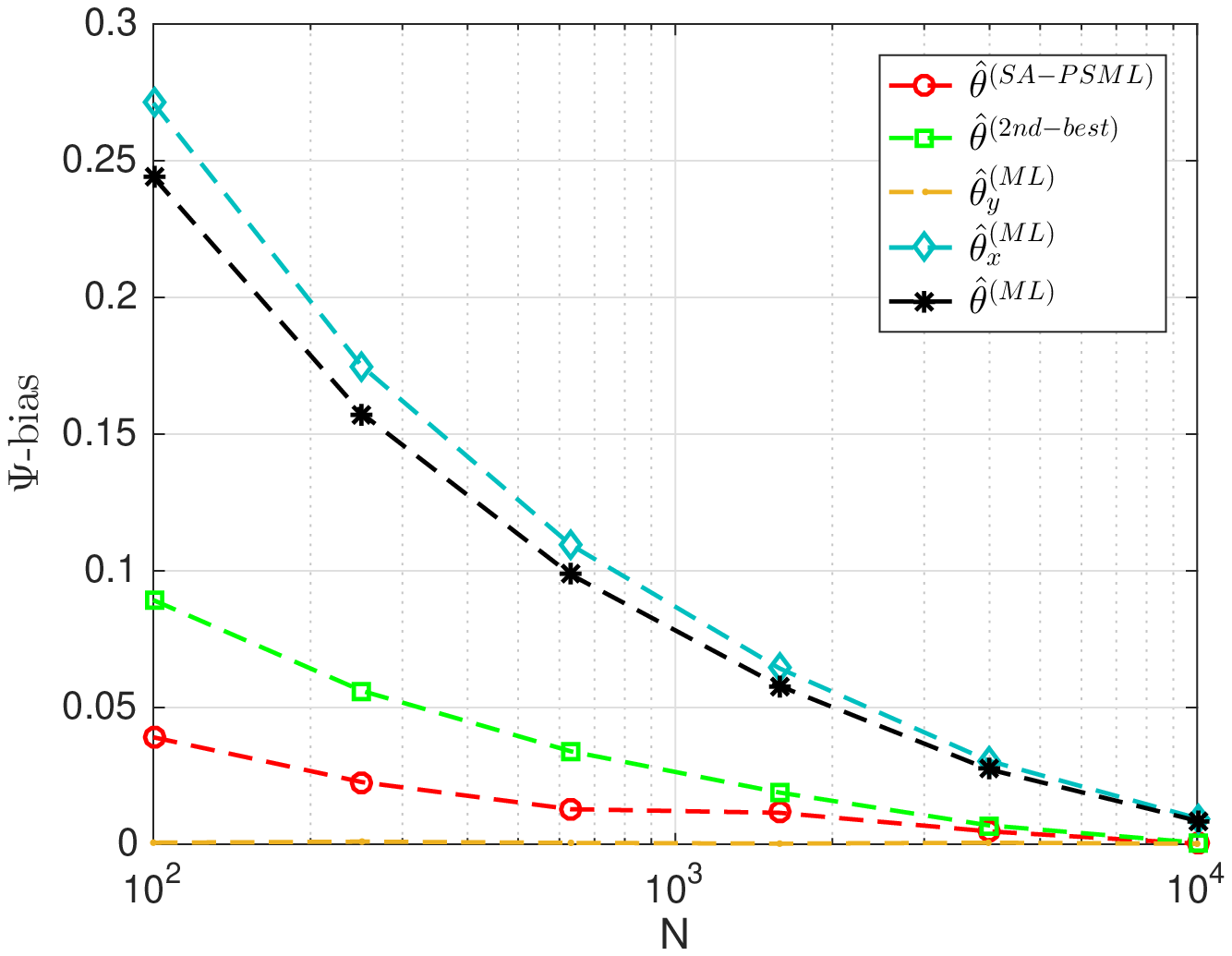}\vspace{-0.2cm}}	 
		\subcaptionbox{	\label{CE_psmse_N} }[\linewidth]
		{\vspace{-0.05cm}\includegraphics[width=6.8cm]{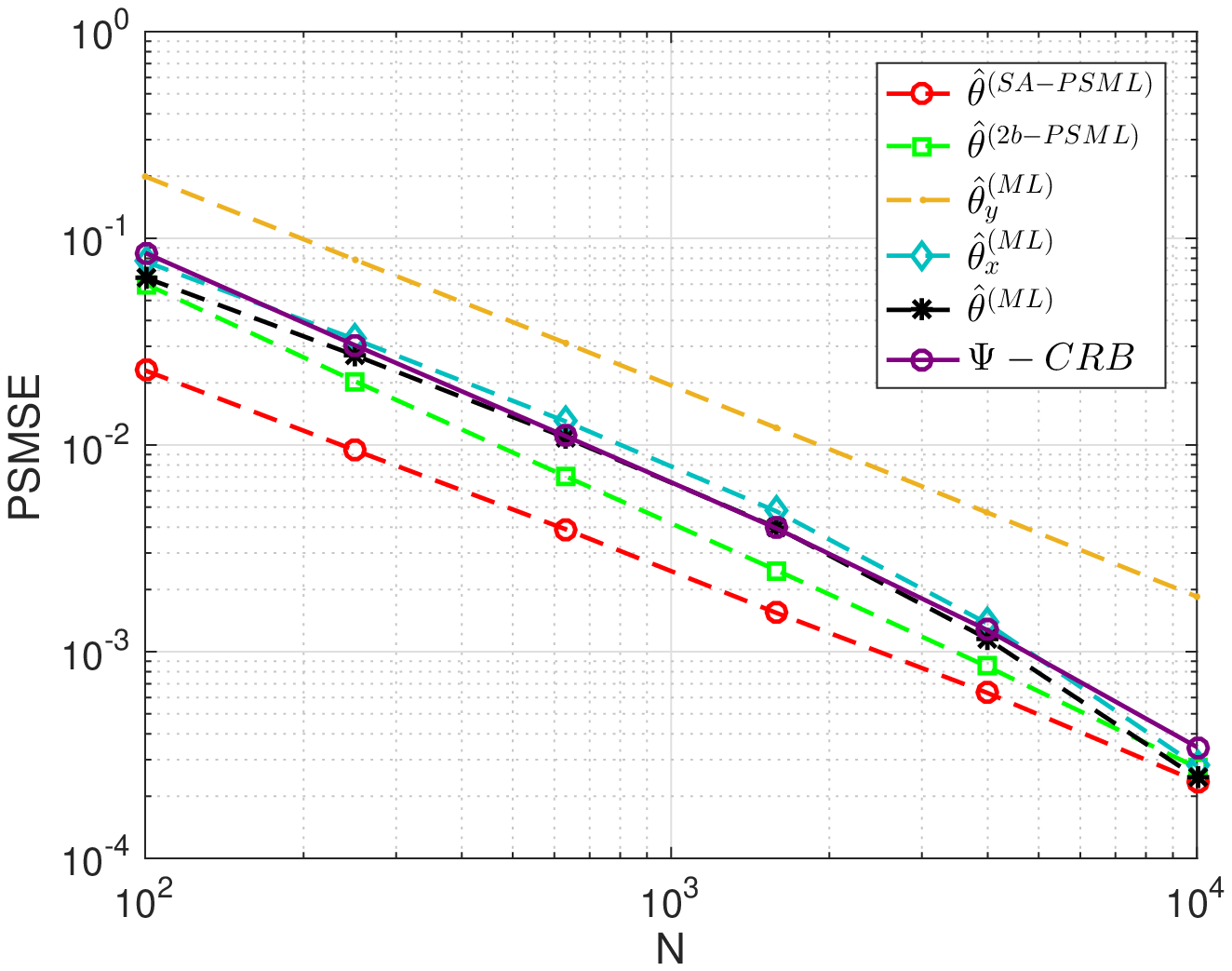}\vspace{-0.2cm}} \vspace{-0.1cm}
		\caption{ Spectrum estimation: The $\Psi$-bias (a) and PSMSE (b) of the SA-PSML and second-best PSML versus the number of observations $N$, compared to split-the-data and the ML estimators.} 
		\vspace{-0.3cm}
		\end{figure}

	In \cref{CE_bias_N,CE_psmse_N} the $\Psi$-bias and PSMSE performance for the spectrum estimation after channel selection problem are presented versus the total number of observations $N$. In this case, $N_{\xvec}=0.9N$, $N_{\yvec}=0.1N$, $M=30$, and $\theta_1=0.95,~ \theta_2=0.96,\theta_3=0.98, \theta_4=\theta_5=\theta_6=3 ,  ~\theta_k=1, ~k=7,\ldots,M$. It can be seen that the proposed PSML methods achieve better performance than the ML estimator in both terms, $\Psi$-bias and PSMSE. The split-the-data estimator, $\tml(\yvec)$, is an $\Psi$-unbiased estimator, but its PSMSE performance is the highest since it uses only part of the observations.
	
		\subsection{Spectrum estimation with ``black-box" selection rule} \label{knn_simulation}
		In this subsection, we demonstrate the robustness of the proposed SA-PSML method for a case where the selection rule is unknown to the estimator. We consider the CR spectrum estimation after channel selection from \Cref{simulations_channel}, where the selection  is based on the k-Nearest Neighbors (kNN) algorithm \cite{cover1967nearest,fukunaga1990introduction}. 
		The Nearest Neighbor decision rule classifies a point as the classification of the nearest point in a set of classified points. The kNN decision rule is an extension, where the decision is based on the majority vote among the $k$ nearest points. The kNN algorithm has been suggested in \cite{jovicic2009cognitive,thilina_Ekram2013machine} in the context of spectrum sensing in CR systems.

\begin{figure}[htb]
			{
			\subcaptionbox{ \label{CE_knn_bias_N} }[\linewidth]
			{\vspace{-0.05cm}\includegraphics[width=6.8cm]{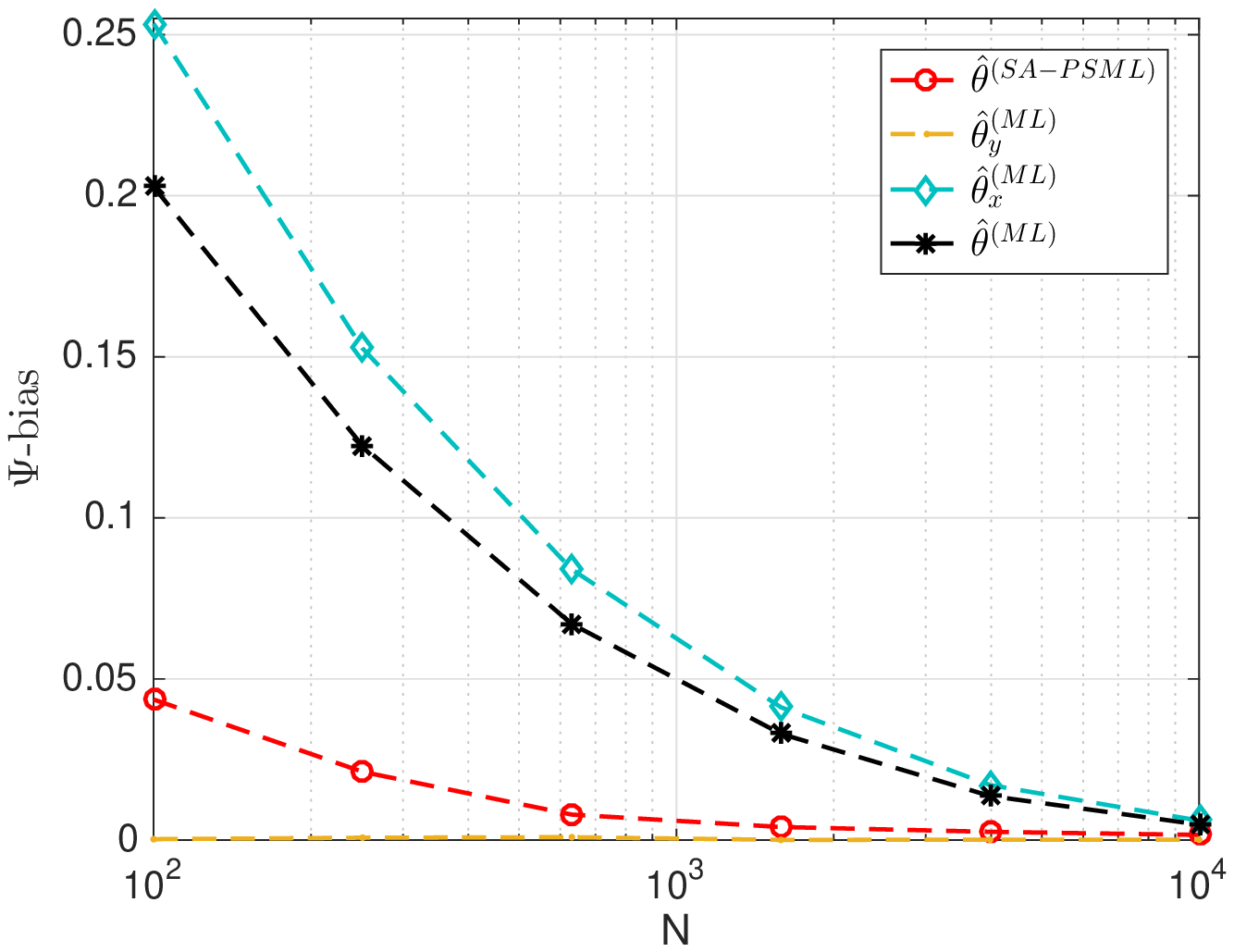}\vspace{-0.2cm}}	 
			\subcaptionbox{	\label{CE_knn_psmse_N} }[\linewidth]
			{\vspace{-0.05cm}\includegraphics[width=6.8cm]{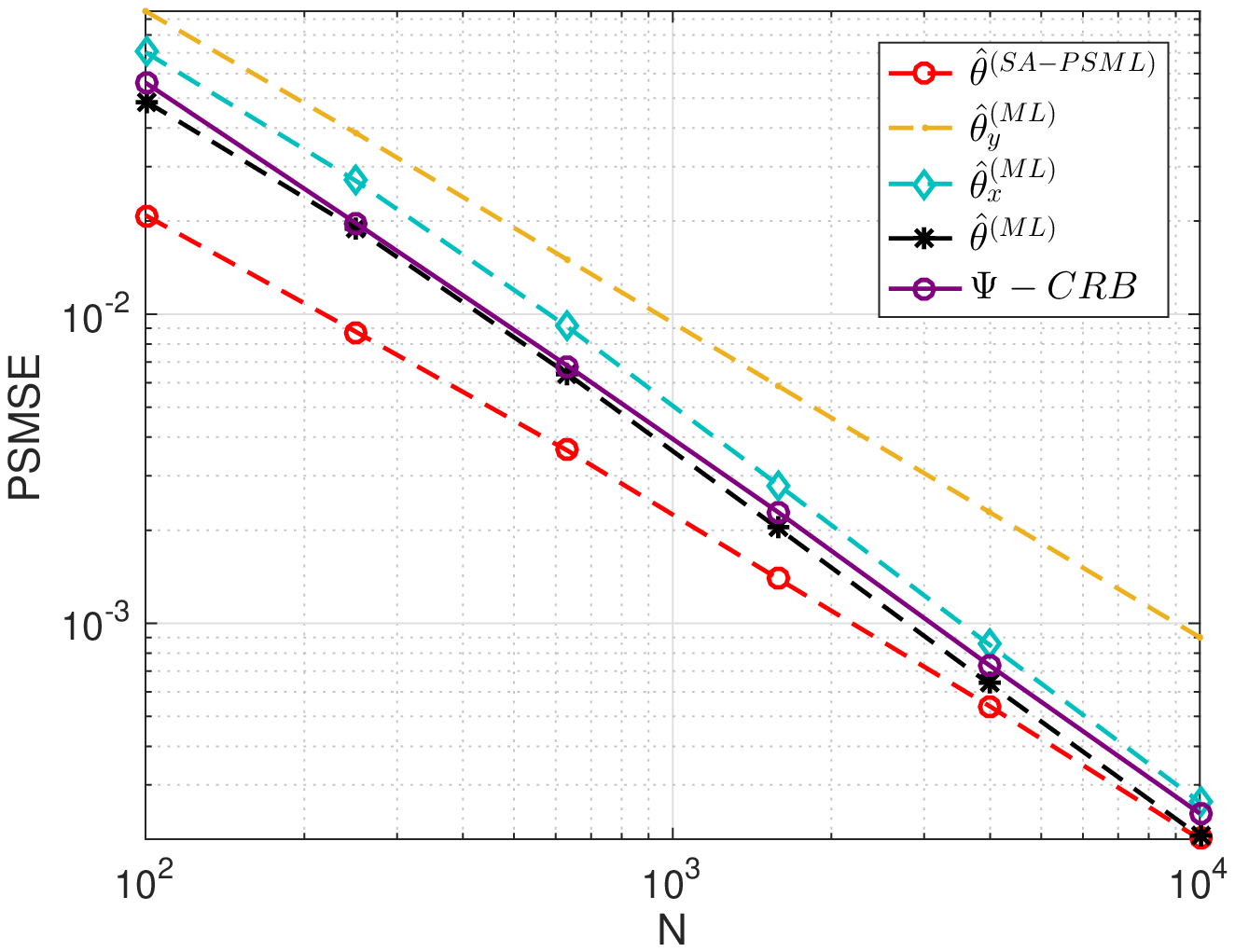}\vspace{-0.2cm}} \vspace{-0.1cm}
			\caption{ Spectrum estimation with a ``black-box" kNN selection rule: The $\Psi$-bias (a) and PSMSE (b) of the SA-PSML, split-the-data, and ML estimators versus the number of observations, $N$.}
		}\vspace{-0.5cm}
		\end{figure}

	Let $\cal X $ be a set of labeled points in $\mathbb{R}^M$, i.e. the ``correct" selection for every point in $\cal X $ is known. For the first stage observation set, $\xvec$, the kNN selection rule by the $k$ nearest vectors in $\cal X$ to $\tml(\xvec)$, is defined in \eqref{sig_ml}. 
		In the following, we assume that the training data-set, $\cal X$, is inaccessible to the estimator directly; therefore, the kNN selection rule can be interpreted as a black-box procedure. {However, we assume that we can generate multiple realizations from the observation model and determine the selection for each realization, to obtain the approximations from \eqref{gvec_indirect}.}
		In \cref{CE_knn_bias_N,CE_knn_psmse_N}, the $\Psi$-bias and PSMSE of the proposed SA-PSML estimator and the ML estimator are shown versus the total number of observations, $N$, where $N_\xvec=0.8N$, $N_\yvec=0.2N$, $\theta_1=0.9, \theta_2=\theta_3=0.95, ~\theta_k=1, ~k=4,\ldots,M$ and $M=25$. 
		It can be seen that although the selection rule is unknown, the SA-PSML estimator have lower $\Psi$-bias and PSMSE than those of the ML estimator. The split-the-data estimator is $\Psi$-biased, but its PSMSE is the highest since it does not use all the observations.
\section{Conclusion} \label{Conclusion}
	In this paper, we derived low-complexity estimation methods for estimation after parameter selection, where the selection of the parameter of interest is data-based and predetermined. First, we established a detailed model and theoretical results, including unbiasedness and the properties of the PSML estimator, for two-stage non-Bayesian estimation after selection. Then, we developed low-complexity methods that take into account the preliminary selection stage and, at the same time, can be implemented for high-dimensional settings. We adopt the MBP algorithm to obtain the PSML estimator. Then, the MBP-PSML estimator is integrated {into} two methods, second-best PSML and SA-PSML, that avoid the need for calculating the probability of selection. In addition, we derive the empirical $\Psi$-CRB, which can be used as a low-complexity performance analysis tool.
	The proposed methods are implemented in a linear Gaussian model, and for spectrum sensing in a CR application. It was shown that these methods achieve significant improvement {in terms of $\Psi$-bias and PSMSE} in comparison to the ML estimator and split-the data estimators and have moderate computational complexity. Topics for future research include derivation of low-complexity methods and bounds for estimation after {\em{model}} selection and for post-detection estimation.
	
\appendices 
\setcounter{section}{-1}
 \section{ Existence of $\Psi$-unbiased estimator} \label[appendix]{proof_psi_unbiased_existence} \vspace{-0.1cm}
 In this appendix we show that for the two-stage model, which satisfies \eqref{pdf_x_y}, if a mean-unbiased estimator of $\thetavec$ based only on the second-stage observation vector, $\yvec$, exists, and under the assumption that $\Omega_\yvec \neq \emptyset$, then, for any selection rule, there exists an $\Psi$-unbiased estimator of $\thetavec$. 
	\begin{lemma} \label{psi_unbiased_existence}
		Let $ \hat{\thvec}(\yvec)$ be an estimator of $\thvec$ based only on the observation set $\yvec$. Assuming that $ \hat{\thvec}(\yvec)$ is a mean-unbiased estimator, {i.e.}
		\be \label{mean_unbiasedness}
		\expec{\thvec}[ \hat{\thvec}(\yvec)-\thvec]= \zerovec,
		\ee
		then $ \hat{\thvec}(\yvec)$ is an $\Psi$-unbiased estimator for the two-stage model.
	\end{lemma}
	\begin{IEEEproof}
	The mean unbiasedness in \eqref{mean_unbiasedness} implies that
\be \label{mean_unbiasedness_m}
	\int_{\Omega^{(m)}_{\yvec}}(\hat{\theta}_m-\theta_m)f_m(\yvec;\thvec) \mathrm{d}\yvec=0,~~\forall m=1,\ldots,M.
\ee
Therefore, for all $ m=1,\ldots,M$ that $\Pr\left(\Psi_\xvec=m;\thvec\right)\neq 0$, we obtain that
\beqna \label{unbiasedness_existence} \begin{aligned}[b]
	&\expec{\thvec}\hspace{-0.1cm}\left[\hat{\theta}_m-\theta_m|\Psi_\xvec=m \right]\\
	&=	\int_{\Am}\int_{\Omega^{(m)}_{\yvec}} (\hat{\theta}_m-\theta_m)  f(\xvec,\yvec|\Psi_\xvec=m;\thvec) \mathrm{d}\yvec\mathrm{d}\xvec\\
	&=\hspace{-0.1cm}	\int_{\Am} \hspace{-0.1cm}  \frac{f(\xvec;\thvec)}{\Pr(\Psi_\xvec=m;\thvec)} \hspace{-0.1cm}\int_{\Omega^{(m)}_{\yvec}}\hspace{-0.1cm}(\hat{\theta}_m-\theta_m)f_m(\yvec;\thvec) \mathrm{d}\yvec\mathrm{d}\xvec \\&=0,
\end{aligned}
\eeqna
where the first equality in \eqref{unbiasedness_existence} is obtained by substituting \eqref{conditioned_likelihood_2stage}, the second equality is obtained by substituting \eqref{pdf_x_y}, and the last equality is obtained by substituting \eqref{mean_unbiasedness_m}. Therefore, the $\Psi$-unbiasedness condition from \eqref{Psi_unbiasedness_term} holds and $ \hat{\thvec}(\yvec)$ is an $\Psi$-unbiased estimator.
\end{IEEEproof}
\FloatBarrier
	

\end{document}